\documentclass[english]{article}
\usepackage[T1]{fontenc}
\usepackage[utf8]{inputenc}
\usepackage{geometry}
\usepackage{color}
\usepackage{babel}
\usepackage{float}
\usepackage{textcomp}
\usepackage{amsmath}
\usepackage{amsthm}
\usepackage{graphicx}
\usepackage{setspace}
\usepackage[unicode=true,pdfusetitle,
 bookmarks=true,bookmarksnumbered=false,bookmarksopen=false,
 breaklinks=false,pdfborder={0 0 1},backref=false,colorlinks=false]
 {hyperref}

\makeatletter
\theoremstyle{plain}
\newtheorem{lyxalgorithm}{\protect\algorithmname}

\makeatother

\providecommand{\algorithmname}{Algorithm}

\begin{document}
\begin{center}
\textbf{\LARGE{}The Democratization of Wealth Management: Hedged Mutual
Fund Blockchain Protocol}{\LARGE\par}
\par\end{center}

\begin{center}
\textbf{\large{}Ravi Kashyap (ravi.kashyap@stern.nyu.edu)}\footnote{\begin{doublespace}
Numerous seminar participants, particularly at a few meetings of the
econometric society and various finance organizations, provided suggestions
to improve the paper. The following individuals have been a constant
source of inputs and encouragement: Dr. Yong Wang, Dr. Isabel Yan,
Dr. Vikas Kakkar, Dr. Fred Kwan, Dr. Costel Daniel Andonie, Dr. Guangwu
Liu, Dr. Jeff Hong, Dr. Humphrey Tung and Dr. Xu Han at the City University
of Hong Kong. The views and opinions expressed in this article, along
with any mistakes, are mine alone and do not necessarily reflect the
official policy or position of either of my affiliations or any other
agency.
\end{doublespace}
}
\par\end{center}

\begin{center}
\textbf{\large{}Estonian Business School / City University of Hong
Kong}{\large\par}
\par\end{center}

\begin{center}
\today
\par\end{center}

\begin{doublespace}
\begin{center}
\textbf{\textcolor{blue}{\href{https://doi.org/10.1016/j.ribaf.2024.102487}{Edited Version: Kashyap, R. (2024).  The Democratization of Wealth Management: Hedged Mutual Fund Blockchain Protocol.  Research in International Business and Finance,  July 2024,  102487. }}}
\par\end{center}
\end{doublespace}

\begin{center}
\par\end{center}

\begin{center}
Journal of Economic Literature Codes: G11 Investment Decisions; D81
Criteria for Decision-Making under Risk and Uncertainty; C32 Time-Series
Models; B23 Econometrics, Quantitative and Mathematical Studies; D8:
Information, Knowledge, and Uncertainty; I31: General Welfare, Well-Being;
O3 Innovation • Research and Development • Technological Change •
Intellectual Property Rights;
\par\end{center}

\begin{center}
intelligence
\par\end{center}

\begin{doublespace}
\begin{center}
\pagebreak{}
\par\end{center}
\end{doublespace}

\begin{center}
\tableofcontents{}\pagebreak{}
\par\end{center}

\begin{doublespace}
\begin{center}
\listoffigures 
\par\end{center}

\begin{center}
\listoftables 
\par\end{center}
\end{doublespace}

\begin{singlespace}
\begin{center}
\pagebreak{}
\par\end{center}
\end{singlespace}
\begin{doublespace}

\section{Abstract}
\end{doublespace}

\begin{doublespace}
\noindent We develop several innovations designed to bring the best
practices of traditional investment funds to the blockchain landscape.
Our innovations combine the superior mechanisms of mutual funds and
hedge funds. Specifically, we illustrate how fund prices can be updated
regularly like mutual funds and performance fees can be charged like
hedge funds. In addition we show how mutually hedged blockchain investment
funds can operate with investor protection schemes - such as high
water marks - and measures to offset trading related slippage costs
when redemptions happen. We provide detailed steps - including mathematical
formulations and instructive pointers - to implement these ideas as
blockchain smart contracts. We discuss how our designs overcome several
blockchain bottlenecks and how we can make smart contracts smarter.
We provide numerical illustrations of several scenarios related to
the mechanisms we have tailored for blockchain implementation.

\noindent The concepts we have developed for blockchain implementation
can also be useful in traditional financial funds to calculate performance
fees in a simplified manner. We highlight two main issues with the
operation of mutual funds and hedge funds and show how blockchain
technology can alleviate those concerns. The ideas developed here
illustrate on one hand, how blockchain can solve many issues faced
by the traditional world and on the other hand, how many innovations
from traditional finance can benefit decentralized finance and speed
its adoption. This becomes an example of symbiosis between decentralized
and traditional finance - bringing these two realms closer and breaking
down barriers between such artificial distinctions - wherein the future
will be about providing better risk adjusted wealth appreciation opportunities
to end customers through secure, reliable, accessible and transparent
services - without getting too caught up about how such services are
being rendered. 
\end{doublespace}

\section{\label{sec:Mutually-Hedging-Investments}Introduction: Mutually Hedging
Decentralized Investment Platforms}

Modern financial investment funds have evolved over a long time period
to their present day form (Goetzmann \& Rouwenhorst 2005; Mallaby
2010). The relatively long progression period of traditional finance
funds - compared to decentralized wealth management - has given rise
to numerous innovative techniques, including but not limited to :
1) to charge fees from customers for the services rendered, 2) mechanisms
to provide liquidity to investors, 3) risk management techniques and
diversified investment products, 4) to ensure fairness in terms of
fees charged and 5) to be able to spread the costs of fund management
over a large user base so that individual costs - and efforts - are
minimized (Silber 1983; Levinthal \& Myatt 1994; Tufano 2003; Matz
\& Neu 2006; Cherkes et al., 2008; Broby 2012; Kavanagh et al., 2014;
Deakin 2015; Lenkey 2015; Cremers et al., 2016; Brown \& Pomerantz
2017; Lo 2017). Clearly there are wider economic implications of a
well developed and robust financial landscape (Minsky 1986; 1990;
De Gregorio \& Guidotti 1995; Levine 1997; Rajan \& Zingales 1998;
Levine 2005; Ahmad et al., 2020).

To simplify our discussion we note that - two broad categories of
investment vehicles - hedge funds and mutual funds, operate differently
in terms of: 1) the investment strategies they choose, 2) the mechanisms
they use to charge fees, 3) the benchmarks chosen to measure their
performance, and 4) the inflow and outflow of investor funds (Fung
\& Hsieh 1999; Liang 1999; Malkiel \& Saha 2005; Eling \& Schuhmacher
2007; Agarwal, Boyson \& Naik 2009; Eling \& Faust 2010; Cici, Gibson
\& Moussawi 2010; End-notes \ref{enu:Open-end-mutual-funds}; \ref{enu:A-hedge-fund}). 

Mutual funds charge a variety of fees to cover their business expenses
(Chordia 1996; Golec 1996; Dellva \& Olson 1998; Elton, Gruber \&
Blake 2003; Gil‐Bazo \& Ruiz‐Verdú 2009; Khorana, Servaes \& Tufano
2009; Cuthbertson, Nitzsche \& O'Sullivan 2010). To facilitate ease
of blockchain implementation we categorize these fees into: 1) deposit
fees (charged to investors when an investment is made), 2) redemption
fees (charged when the investor makes a withdrawal request) and 3)
management fees which depend on the assets under management (AUM:
End-note \ref{enu:Finance-AUM}). 

With some abuse of terminology, AUM could also be termed the total
value locked (TVL: End-note \ref{enu:TVL}) using decentralized finance
(DeFi: Zetzsche, Arner \& Buckley 2020; Werner, et al., 2021; Grassi
et al., 2022; End-notes \ref{enu:Decentralized-finance}; \ref{enu:Types-Yield-Enhancement-Services})
lingo. Management fees do not directly depend on the growth in the
value of the investments, that is on the appreciation of the asset
prices. Mutual funds typically hold liquid investments and investors
can buy or sell mutual fund shares (a unit) periodically (usually
daily) based on the price of the shares (also know as Net Asset Value,
NAV: Penman 1970; End-note \ref{EN:Net-Asset-Value}) which is also
updated on a corresponding time interval (usually also a daily basis). 

Hedge funds have a key distinction since they charge some form of
performance fees based explicitly on the returns they generate for
investors (Agarwal, Daniel \& Naik 2009; Ben-David, Birru \& Rossi
2020). Performance fees - which are intrinsically tied to the growth
in the value of investments that investors receive - are usually collected
separately from management fees. Hedge funds have restrictions on
investor redemptions and the NAV is not publicly made available like
mutual funds (Aragon 2007; Cumming \& Dai 2009; Hong 2014; Aiken,
Clifford \& Ellis 2015). 

Two main issues pertaining to modern investment funds are that: 1)
not all investment opportunities are available for all investors (Asness,
Krail \& Liew 2001; Brooks \& Kat 2002; Stulz 2007), and 2) the holdings
of the funds are not very transparent to outsiders (Anson 2002; Haslem
2004; 2007; Hedges 2005; Prat 2005; Goltz \& Schröder 2010; Aggarwal
\& Jorion 2012; Aragon, Hertzel \& Shi 2013; Agarwal,Vashishtha \&
Venkatachalam 2018). Both these issues can be solved, to a great extent,
by utilizing blockchain technology. In this article we provide several
innovations that will enable investment funds to operate using decentralized
ledgers (Nakamoto 2008; Di Pierro 2017).

There are pros and cons to the operational dynamics of both hedge
funds and mutual funds. We cherry pick the superior mechanisms from
both types of funds - mutual funds and hedge funds - and take them
to the blockchain investment space. 

\subsection{\label{subsec:Blockchain-Hedged-Mutual}Blockchain Investment Funds:
Challenges and Solutions}

There are several works that describe the application of blockchain
technology to fund management and also to other aspects of financial
intermediation (Manda et al., 2018; 2023; Patel, Migliavacca \& Oriani
2022; Ren et al., 2023). These pioneering works provide a conceptual
overview regarding the procedures pertaining to fund management that
can be transferred to decentralized ledger based solutions. Streamlining
some of the fund management processes onto blockchain can bring about
significant operational efficiencies - that will reduce costs - and
improved security due to the inherent cryptographic features of decentralized
technologies, while making data publicly available for both reporting
and decision making. (Fiergbor 2018; Ciriello 2021; Srivastava 2023)
provide detailed examples pertaining to index funds - that is mutual
funds and exchange traded funds (End-note \ref{enu:An-index-fund}).

It is important to clarify that the existing works on blockchain fund
management focus on providing managerial insights and high level overviews
on how to approach the nuances of decentralized technology that can
be applied to financial wealth management. In contrast, we have designed
a complete fund management blockchain protocol - including detailed
mathematical formulations and pointers - that is ready for software
implementation. 

We have also created several innovations - related to investor protection
and seamlessly levying fees - applicable to blockchain fund management
in terms of adapting best practices from traditional finance to the
blockchain realm. Initially, we describe the new techniques we have
created at a high level, so that they are easy to understand for a
wide audience. Later sections give granular steps - so that they can
help participants comprehend the complexities involved and also to
be able to build the necessary infrastructure. The outline of the
various topics is given later in Section (\ref{subsec:Outline-of-the}).

Our novel techniques combine the superior mechanisms of investment
funds - mutual funds and hedge funds - while provide detailed steps
to implement them as blockchain smart contracts (Buterin 2014; Macrinici
et al., 2017; Mohanta et al., 2018; Hu et al., 2021; Negara et al.,
2021; Vacca et al., 2021; Sharma et al., 2023; End-note \ref{enu:A-smart-contract}).
We discuss how our designs overcome several blockchain bottlenecks
and how we can make smart contracts smarter. Specifically, we illustrate
how fund prices can be updated regularly like mutual funds and performance
fees can be charged like hedge funds. In addition, we show how mutually
hedged blockchain investment funds can operate with investor protection
measures such as high water marks and methodologies to offset trading
related slippage costs when redemptions happen. We provide numerical
illustrations of several scenarios related to the mechanisms we have
tailored for blockchain implementation.

\begin{doublespace}
\noindent The concepts we have developed for blockchain implementation
can also be useful in traditional financial funds to calculate performance
fees in a simplified manner. We show how blockchain technology can
alleviate concerns regarding two issues we have highlighted with the
operation of mutual funds and hedge funds. The ideas developed here
illustrate, on one hand, how blockchain can solve many issues faced
by the traditional world, and on the other hand, how many innovations
from traditional finance - tried and tested over time - can benefit
decentralized finance and speed its adoption. This becomes an example
of symbiosis between decentralized and traditional finance - bringing
these two realms closer and breaking down barriers between such artificial
distinctions - wherein the future will be about providing better risk
adjusted wealth appreciation opportunities to end-customers through
secure, reliable, accessible and transparent services - without getting
too caught up about how such services are being rendered. 
\end{doublespace}

Below we describe the challenges - and the corresponding solutions
we have pioneered - to bring traditional financial investment fund
operational models - and best practices - to the Blockchain landscape.
We wish to utilize the technological developments that have happened
in the many decades since hedge funds became popular and to also combine
several enhancements related to the blockchain decentralization paradigm.
We lay down the techniques for an investment platform that keeps all
fund assets entirely on the blockchain platform at all times. Kashyap
(2021) has a detailed discussion of other related topics that are
aimed at making blockchain investing secure and less risky. The specific
issue of maintaining assets on-chain - in a highly secure manner -
is also discussed in one of the earlier sections of Kashyap (2021).

\subsection{\label{subsec:Fees-for-Protection}Fees for Protection with Appreciation:
Quid Pro Quo}

The community driven spirit of DeFi makes it essential to have some
form of profit sharing, wherein part of any excess fees generated
is given back to long term investors of the project (Singh \& Kim
2019; Wang et al., 2019; El Faqir et al., 2020; Wu et al., 2021; Zheng
\& Boh 2021; Ballandies et al., 2022; Liu et al., 2022; Kitzler et
al., 2023; Saurabh et al., 2023). Fueling the decentralized ethos
of profit sharing are many grievances related to high fees in the
traditional financial sector, with counter arguments being provided
about competition between funds restricting fees to reasonable levels
(Golec 1996; Coates \& Hubbard 2007; Khorana et al., 2009; Malkiel
2013; Philippon 2015; Feldman et al., 2020). With blockchain technology
it becomes easy to assess profits, fees and related metrics - due
to the transparency of operations across all types of decentralized
investment vehicles - given that making information publicly available
is a foundational principle built right into the genes.

While the importance of sharing the proceeds - profits from operations
– with all the participants cannot be emphasized enough, it is equally
- if not more - important to ensure that the community understands
the need to have funds set aside for a rainy day and the necessity
of being able to invest in projects that can prepare the community
for a better tomorrow. Most protocols that wish to follow the mechanisms
designed here need to emphasize to investors - community - that the
performance fees are to ensure that there are enough funds to sustain
operating expenses and fund future growth plans. The performance fees
will ensure that investments remains stable across market cycles -
and continue to grow - rewarding investors over the long term. Additional
performance fees, above a threshold, will be returned back to the
community (Kashyap 2021). 

An extremely popular investor protection mechanism, in the traditional
finance world, is the idea of a high water mark (HWM: Goetzmann, et
al., 2003; Guasoni \& Obłój 2016; End-note \ref{enu:High-water-mark}).
The simple summary of this concept is that performance fees are charged
only when investors are entitled to a profit derived from their original
principal. This is, perhaps, best clarified with a simple numerical
illustration. 

For example, let us say an investor deposits 10,000 USD. After some
time, the invested amount grows to 14,000 USD, at which a high water
mark is established. The profit in this case is 4,000 USD. A part
of this profit is taken as performance fees. After this, if the value
of the investment goes down to say 12,000 no performance fees are
charged until the value of investment climbs back above 14,000, the
high water mark. The bottomline is that unless a tangible wealth increase
is generated for every investor - at a holistic level - no performance
fees are paid. This creates a strong incentive for the team to produce
solid - and quantifiably measurable - returns for the investors.

This simple scenario can get extremely complicated when there are
multiple investors who deposit at different levels of the fund price.
Tracking all this in a smart contract - with the current state of
blockchain technology - is extremely hard and can be deemed almost
impossible (Giancaspro 2017; Wang, et al., 2018; Huang et a., 2019;
Zou, et al., 2019; Sayeed et al., 2020; Zheng, et al., 2020; Tolmach
et al., 2021; Kannengießer et al., 2021). To be able to accommodate
these complexities we have found a novel solution that works elegantly
- is rather straightforward to implement as a smart contract - and
provides the same level of protection to every single investor. Our
solution - which utilizes weighed average calculations - is also mathematically
identical, in terms of aggregate fees and proceeds, to what investment
funds in the traditional world have been doing for decades. 

In addition, the detailed algorithm we have developed - to create
a fund management protocol - ensures that other investor protection
schemes are incorporated. In particular, we have created techniques
to maintain fairness while investors enter or exit the fund. Limits
- maximum amounts - are set on the total amounts that can enter or
leave the fund during any time interval. The overall objective is
to reward investor loyalty and their preference for staying longer
with the fund, which will benefit all participants - the entire community. 

Incentivizing investor allegiance is done by letting investor money
into the fund using a queue - on a first come first served basis (Kruse
1984; Andrew \& Herbert 2015; End-note \ref{enu:In-computing-FIFO}).
Exiting the fund is done so that at each rebalancing interval, the
maximum redemption amount for the entire fund is allocated across
all investors wanting to take out their money on a proportionate basis.
That is, if a full redemption cannot be made, then everyone gets the
same percentage of their withdraw amount satisfied. This is to be
take care of the possibility that market runs - and other fund liquidation
issues - are avoided during panic driven sell offs (Renshaw 1984;
Kitamura 2010; Kleinnijenhuis et al., 2013; Chen \& Huang 2018; Huynh
\& Xia 2023). 

Also, the buying and selling of assets affects asset prices and is
known as slippage (Bertsimas \& Lo 1998; Kashyap 2020; End-note \ref{enu:In-financial-markets,}).
We distribute the price appreciation, and the depreciation, when investor
money enters, or leaves, the fund across all participants using the
treasury - to finance shortfalls, or to collect surplus amounts -
so that no one benefits from large deposits, or gets penalized when
large sells are made.

\subsection{\label{subsec:Outline-of-the}Outline of the Sections Arranged Inline}

Section (\ref{sec:Mutually-Hedging-Investments}) - which we have
already seen - provides an introductory overview of modern investment
funds, the motivations for bringing such investment vehicles to blockchain
and the innovations we are creating to apply traditional finance principles
to the blockchain realm. Section (\ref{subsec:Blockchain-Hedged-Mutual})
describes specifically how our contributions add to the various efforts
being undertaken to bring traditional wealth management to blockchain.
Section (\ref{subsec:Fees-for-Protection}) discuss the intuitions
behind the innovations we are adopting in the blockchain wealth management
realm to protect investors and charge fees. Later sections provide
detailed mathematical steps and technical pointers.

Section (\ref{sec:Sequence-of-Steps}) gives an algorithm to accept
deposits and withdrawal requests from investors, after calculating
the price of fund shares or tokens. Section (\ref{sec:Performance-Fee-HWM})
is a discussion of a novel technique to charge performance fees, and
maintain high water marks, despite the limitations of smart contracts.
Section (\ref{subsec:Blockchain-Bottlenecks-and}) considers the challenges
of performance fee calculations in a decentralized environment. Section
(\ref{subsec:High-Water-Markets}) outlines issues, for charging fees,
that funds face when the NAV trends lower - due to falling market
prices - than the fund water mark. Section (\ref{subsec:Performance-Fees-Across-Aggregate})
outlines our solutions to address the problem in Section (\ref{subsec:High-Water-Markets})
tailored to perform seamlessly on blockchain networks. Section (\ref{subsec:Performance-Fees-Across-Separate})
is a discussion of how fee levies can happen in the absence of our
solutions in Section (\ref{subsec:Performance-Fees-Across-Aggregate})
and to help readers understand the significance of the techniques
we have developed.

Section (\ref{sec:Numerical-Results}) explains the numerical results
we have obtained, which illustrate how our innovations compare to
existing wealth management techniques. Section (\ref{sec:Fund-Flow-Flow})
has the flow charts related to the material discussed in Sections
(\ref{sec:Sequence-of-Steps}; \ref{sec:Performance-Fee-HWM}). The
diagram in Section (\ref{sec:Fund-Flow-Flow}) is given for completion
and for helping readers obtain a better understanding of the concepts
involved. Sections (\ref{sec:Areas-for-Further}; \ref{sec:Conclusion})
suggest further avenues for improvement and the conclusions respectively. 

\section{\label{sec:Sequence-of-Steps}Periodic Blockchain Fund Management
Methods: Sequences of Steps}
\begin{lyxalgorithm}
\label{alg:Sequence-Steps-algorithm}The following algorithm captures
the sequence of steps, that need to be carried out at periodic intervals,
to accomplish secure fund management entirely on a blockchain environment.
The interaction of the various processes described in the following
points facilitate secure movement of assets, rigorous risk management
and rebalancing the portfolio to adhere to asset weights that match
risk and return objectives.
\end{lyxalgorithm}
\textbf{\textit{For some steps, there might be further sub steps,
making these sub steps a sequence of steps within a sequence of steps
and as a consequence the entire thing becomes ``Sequences of Steps''.}}
\begin{itemize}
\item Figure (\ref{fig:Sequences-of-Steps}) in Section (\ref{sec:Fund-Flow-Flow})
has the flow chart corresponding to the steps given in Algorithm (\ref{alg:Sequence-Steps-algorithm}).
\item Figures (\ref{fig:Sequences-of-Steps-Input-Variables}; \ref{fig:Sequences-of-Steps-System-Variables};
\ref{fig:Sequences-of-Steps-Deposit-Withdraw-Accept}) in Section
(\ref{sec:Numerical-Results}) give several scenarios pertaining to
the steps and calculations given here. The figures illustrate different
variables corresponding to inputs and calculated variables, which
should help in monitoring how the system is performing.
\end{itemize}
\begin{enumerate}
\item \label{enu:Calculate-Total-Fund}Calculate Total Fund Value across
all chains. This is the sum of each asset quantity multiplied by the
corresponding asset price. For vaults and liquidity pools, this will
be the dollar value invested in that investment opportunity. Note
that we specify ``across all chains'' since assets will be held
on multiple chains with funds invested accordingly. We call this fund
Alpha going forward for ease of reference.
\item \label{enu:Check-Existing-Number}Check Existing Number of Tokens
for the Alpha investment fund (across all chains). Tokens represent
the number of units, or the quantity of any asset. They are the same
as shares of stocks in the traditional financial world.
\item \label{enu:Calcualte-new-NAV}Calculate new NAV - or Alpha Price or
Fund Price - based on Step (\ref{enu:Calculate-Total-Fund}) and Step
(\ref{enu:Check-Existing-Number}). The Alpha price will be the same
for the entire fund and across all the chains.
\item \label{enu:Take-out-Performance-Fees}Calculate Performance Fees and
Management Fees based on new NAV from Step (\ref{enu:Calcualte-new-NAV})
on each chain separately. The calculation of performance fees has
several steps of its own and is discussed in detail in Section (\ref{sec:Performance-Fee-HWM}).
\item \label{enu:Issue-new-Alpha}Issue new Alpha Tokens, as necessary for
performance fees and management fees, for Step (\ref{enu:Take-out-Performance-Fees})
on each chain separately.
\item \label{enu:Adjust-NAV-based}Adjust NAV based on Alpha Tokens issued
for performance fees and management fees in Step (\ref{enu:Issue-new-Alpha})
across all chain separately.
\item \label{enu:Rebalance-net-new}Aggregate the deposit amounts and withdrawal
amounts so that the net total amount received for investment, or for
redemption, is less than the corresponding maximum amount. 
\begin{itemize}
\item Note that deposit requests are made in dollar denominations and withdraw
requests are based on the number of tokens. 
\item The maximum amounts for deposit, or withdraw, for any rebalancing
event is to ensure that there is a limit to how much funds can be
taken out of, or invested into, the fund during any given time interval. 
\item Also, note that the maximum deposit amount and maximum withdrawal
amount per rebalance event could be different. 
\item We have to look at two cases depending on whether money will flow
into the fund, or out of the fund, during this rebalancing sequence
of events. That is we set flags $NETDPSTIND_{t}$ or $NETWDRWIND_{t}$,
which will indicate whether the net amount is positive (net deposit)
or negative (net withdrawal). 
\begin{equation}
NETDPSTIND_{t}=\begin{cases}
1, & TOTALDPSTUSD_{t}\geq\left(TOTALWDRWTOKENS_{t}*ALPHAPRICE_{t}\right)\\
0, & TOTALDPSTUSD_{t}<\left(TOTALWDRWTOKENS_{t}*ALPHAPRICE_{t}\right)
\end{cases}
\end{equation}
\begin{equation}
NETWDRWIND_{t}=\lnot\left(NETDEPOSIT_{t}\right)
\end{equation}
\end{itemize}
Here, 
\begin{equation}
TOTALWDRWTOKENS_{t}=\sum_{i=1}^{W_{t}}WDRWREQUEST_{it}
\end{equation}
\begin{equation}
TOTALDPSTUSD_{t}=\sum_{i=1}^{D_{t}}DPSTREQUEST_{it}
\end{equation}

\begin{itemize}
\item $W_{t}$ is the total number of withdrawals requested at time $t$,
before this particular rebalance sequence of events has started. It
is to be understood henceforth, unless otherwise stated, that the
aggregations include items since the last rebalance event has completed.
\item $D_{t}$ is the total number of deposit requests at time $t$, before
this particular rebalance sequence of events has started. 
\item $ALPHAPRICE_{t}$ is the adjusted Alpha price - or fund price - calculated
at Step (\ref{enu:Adjust-NAV-based}). 
\item $WDRWREQUEST_{it}$ is the number of tokens requested for withdraw
by request $i$ at time $t$.
\item $DPSTREQUEST_{it}$ is the dollar amount requested for deposit by
request $i$ at time $t$. 
\item $TOTALWDRWTOKENS_{t}$ is the total number of Alpha tokens requested
for withdrawal at time $t$ before this particular rebalance sequence
of events started. 
\item $TOTALDPSTUSD_{t}$ is the total amount in USD requested for deposit
at time $t$ before this particular rebalance sequence of events started. 
\item We also calculate the net amount per rebalance event, $NETAMOUNTEVENT_{t}$
as follows, 
\begin{align}
NETAMOUNTEVENT_{t} & =\left(NETDPSTIND_{t}\right)*\\
 & \min\left\{ NETDPSTORWDRW_{t},MAXDPSTUSD_{t}\right\} \\
 & -\left(NETWDRWIND_{t}\right)*\\
 & \min\left\{ \left|NETDPSTORWDRW_{t}\right|,MAXWDRWUSD_{t}\right\} 
\end{align}
\begin{equation}
NETDPSTORWDRW_{t}=TOTALDPSTUSD_{t}-\left(TOTALWDRWTOKENS_{t}*ALPHAPRICE_{t}\right)
\end{equation}
\item $MAXWDRWUSD_{t}$ is the maximum amount in USD that can be accepted
for redemptions at time $t$ during any one rebalancing sequence or
event. 
\item $MAXDPSTUSD_{t}$ is the maximum amount in USD that can be accepted
for investment at time $t$ during any one rebalancing event.
\end{itemize}
\begin{enumerate}
\item Case: $NETDPSTIND_{t}=1$. The deposit amounts accepted for investment
are aggregated based on the first in and first out principle depending
on the time stamp the deposit request is made. This is given by the
formula,
\begin{align}
DPSTACCEPT_{it} & =\begin{cases}
DPSTREQUEST_{it}, & i\leq DA_{t}\\
DPSTREQUEST_{it}+\sum_{i=1}^{DA_{t}}DPSTREQUEST_{it} & i>DA_{t}\\
-\left|NETAMOUNTEVENT_{t}\right|\\
-\left|TOTALWDRWTOKENS_{t}*ALPHAPRICE_{t}\right|,
\end{cases}
\end{align}

\begin{itemize}
\item $DPSTACCEPT_{it}$ is the USD amount that will be accepted for deposit
from the total amount requested for deposit from request $i$, $DPSTREQUEST_{it}$
at time $t$. 
\item $DA_{t}$ is the total number of deposit requests at time $t$ that
satisfy the below conditions.
\begin{align}
\sum_{i=1}^{DA_{t}}DPSTREQUEST_{it} & <\left|NETAMOUNTEVENT_{t}\right|\\
 & +\left|TOTALWDRWTOKENS_{t}*ALPHAPRICE_{t}\right|
\end{align}
\begin{align}
\sum_{i=1}^{DA_{t}+1}DPSTREQUEST_{it} & \geq\left|NETAMOUNTEVENT_{t}\right|\\
 & +\left|TOTALWDRWTOKENS_{t}*ALPHAPRICE_{t}\right|
\end{align}
\item $DPSTACCEPTRATIO_{t}$ is the deposit accept ratio which gives the
percentage of the total requested deposit amount accepted for investment
into the fund at this rebalance event, that is at time $t$. It is
a helpful variable for monitoring the system performance given by
the formulation,
\begin{align}
DPSTACCEPTRATIO_{t} & =\min\left(\left\{ \frac{\left|NETAMOUNTEVENT_{t}\right|}{TOTALDPSTUSD_{t}}\right.\right.\\
+ & \left.\left.\frac{\left|TOTALWDRWTOKENS_{t}*ALPHAPRICE_{t}\right|}{TOTALDPSTUSD_{t}}\right\} ,1\right)
\end{align}
\end{itemize}
\begin{enumerate}
\item Issue (mint) new Alpha Tokens to fulfill the deposit requests that
were accepted for investment and remove (burn) Alpha tokens to fulfill
the withdraw requests that were accepted for redemption as necessary
using the NAV from Step (\ref{enu:Adjust-NAV-based}).
\item Rebalance the portfolio using the net amount for deployment, $NETAMOUNTEVENT_{t}$,
based on corresponding portfolio asset weights using the rebalancing
algorithm from Kashyap (2021). 
\item Update the Alpha price again using the first three Steps (\ref{enu:Calculate-Total-Fund};
\ref{enu:Check-Existing-Number}; \ref{enu:Calcualte-new-NAV}). Note
that the total number of tokens now includes the tokens issued for
the amount, $NETAMOUNTEVENT_{t}$, that just got invested.
\end{enumerate}
\item Case: $NETWDRWIND_{t}=1$. The withdrawal tokens accepted for redemption
are aggregated based on the total number of requests with each request
being filled a certain percentage of the requested tokens depending
on the available capacity. This is given by the formula,
\begin{align}
WDRWACCEPT_{it} & =\min\left[\left\{ \frac{\left|NETAMOUNTEVENT_{t}\right|+\left|TOTALDPSTUSD_{t}\right|}{\left(TOTALWDRWTOKENS_{t}*ALPHAPRICE_{t}\right)}\right\} ,1\right]\\
 & *WDRWREQUEST_{it}
\end{align}

\begin{itemize}
\item $WDRWACCEPT_{it}$ is the number of tokens that will be accepted for
redemption from the total tokens requested for withdraw from request
$i$, $WDRWREQUEST_{it}$ at time $t$. 
\item $WDRWACCEPTRATIO_{t}$ is the withdraw accept ratio which gives the
percentage of the total requested withdraw quantity accepted for redemption
at this rebalance event, that is at time $t$. It is a helpful variable
for monitoring the system performance given by the formulation,
\begin{align}
WDRWACCEPTRATIO_{t} & =\min\left[\left\{ \frac{\left|NETAMOUNTEVENT_{t}\right|+\left|TOTALDPSTUSD_{t}\right|}{\left(TOTALWDRWTOKENS_{t}*ALPHAPRICE_{t}\right)}\right\} ,1\right]
\end{align}
\end{itemize}
\begin{enumerate}
\item Rebalance the portfolio using the net amount for deployment, $NETAMOUNTEVENT_{t}$,
based on the corresponding portfolio asset weights, using the rebalancing
algorithm from Kashyap (2021). 
\begin{itemize}
\item The proceeds from this rebalancing will be denoted as $RBLNCPROCEEDS_{t}$
which can be different from $NETAMOUNTEVENT_{t}$ due to market impact
or slippage (Kashyap 2020). $RBLNCPROCEEDS_{t}$ are the proceeds
obtained from trading - denominated in USD - at this rebalancing event. 
\end{itemize}
\item If the percentage difference between the $NETAMOUNTEVENT_{t}$ and
$RBLNCPROCEEDS_{t}$ is higher than the withdrawal slippage tolerance,
$WDRWSLIPTOLERANCE_{t}$ as shown below, PLEASE STOP AND DO NOT GO
TO NEXT STEP. $WDRWSLIPTOLERANCE_{t}$ is a tolerance level we can
set in percentage that shows how much withdraw related trading slippage
we are willing to accept at this rebalancing event.
\begin{itemize}
\item We might have to manually intervene and decide what to do next. 
\item If the assets were not sold, then we have to go back to Step (\ref{enu:Calculate-Total-Fund})
and restart all over again at a more favorable market time. 
\item If the assets are not sold, then we might have to revert back the
performance or management fees. 
\item If the assets were actually sold for a very low price - or partially
sold - we have to again manually intervene and decide how to proceed.
\begin{equation}
\frac{\left|RBLNCPROCEEDS_{t}\right|-\left|NETAMOUNTEVENT_{t}\right|}{\left|NETAMOUNTEVENT_{t}\right|}<\left(-1\right)*WDRWSLIPTOLERANCE_{t}
\end{equation}
\end{itemize}
\item \label{enu:Slippage-Difference}We consider the difference, $RBLNCSLIPPAGE_{t}$
between $NETAMOUNTEVENT_{t}$ and $RBLNCPROCEEDS_{t}$, and handle
the two cases that arise accordingly as per below (Sub-Steps \ref{enu:If-Slippage-Positive};
\ref{enu:If-Slippage-Negative}),
\begin{equation}
RBLNCSLIPPAGE_{t}=\left|RBLNCPROCEEDS_{t}\right|-\left|NETAMOUNTEVENT_{t}\right|
\end{equation}
\item \label{enu:If-Slippage-Positive}If $RBLNCSLIPPAGE_{t}>0$ is positive,
send this amount to the treasury and burn, or remove, Alpha tokens
equal to this amount from the treasury using the NAV, or Alpha Price,
calculated in Step (\ref{enu:Adjust-NAV-based}). 
\begin{itemize}
\item If the treasury does not have enough Alpha tokens then send to the
treasury the amount that is covered by the Alpha tokens in the treasury,
burn the corresponding number of Alpha tokens and reinvest the rest
of the money into the fund in the next rebalance sequence.
\end{itemize}
\item \label{enu:If-Slippage-Negative}If $RBLNCSLIPPAGE_{t}<0$ is negative,
take stable coins (Ante, Fiedler \& Strehle 2021; End-note \ref{enu:Stablecoin})
from the treasury equal to this amount and issue new Alpha tokens
equal to this amount using the NAV, or Alpha Price, calculated in
Step (\ref{enu:Adjust-NAV-based}). Send the tokens to the treasury. 
\begin{itemize}
\item If the treasury does not have enough stable coins PLEASE STOP.
\end{itemize}
\item \label{enu:Slippage-Alpha-Update}If we calculate the Alpha price
at this time using the first three Steps (\ref{enu:Calculate-Total-Fund};
\ref{enu:Check-Existing-Number}; \ref{enu:Calcualte-new-NAV}), it
will be different from the Alpha price calculated earlier in Step
(\ref{enu:Adjust-NAV-based}). 
\begin{itemize}
\item This is because the total number of tokens now includes the tokens
burned, or minted, for the amount, $RBLNCSLIPPAGE_{t}$, that were
removed, or issued, in Sub-Steps (\ref{enu:If-Slippage-Positive};
\ref{enu:If-Slippage-Negative}). 
\item This shows the effect of the slippage on the Alpha price, but this
new Alpha price cannot be used to fulfill withdraw requests since
with this changed price, we might not have enough stable tokens to
satisfy the redemptions. 
\item This Alpha price is for reference alone to show the impact of the
slippage, but it can be used to update the Alpha price after the withdraw
requests (also deposit requests) are fulfilled.
\end{itemize}
\item Remove (burn) Alpha tokens to fulfill the withdraw requests that were
accepted for redemption and issue (mint) new Alpha Tokens to fulfill
the deposit requests that were accepted for investment as necessary
using the NAV, or Alpha Price, calculated in Step (\ref{enu:Adjust-NAV-based}). 
\begin{itemize}
\item Note that the Alpha price in Step (\ref{enu:Adjust-NAV-based}) can
be quite different from the Alpha price after the adjustments from
Step (\ref{enu:Slippage-Alpha-Update}).
\end{itemize}
\end{enumerate}
\end{enumerate}
\begin{itemize}
\item Note that when we mint new Alpha tokens we accept stable tokens (or
other currencies at later stage, Kashyap 2021) in exchange for the
Alpha tokens. Similarly when we burn Alpha tokens we send stable tokens
back to the investor.
\item We prefer the method described above (Sub-Steps \ref{enu:Slippage-Difference};
\ref{enu:If-Slippage-Positive}; \ref{enu:If-Slippage-Negative};
\ref{enu:Slippage-Alpha-Update}). But another option for the withdrawal
case is to transfer cash between chains and update the Alpha price
using the following formula,
\begin{equation}
ALPHAPRICE_{t}=\frac{\left|RBLNCPROCEEDS_{t}\right|+\left|TOTALDPSTUSD_{t}\right|}{\left(\sum_{i=1}^{W_{t}}WDRWACCEPT_{it}\right)}\label{eq:Alpha-Price-Redemption}
\end{equation}
\end{itemize}
\end{enumerate}

\section{\label{sec:Performance-Fee-HWM}Automated Performance Fee Levy Techniques}

In any wealth management fund, to calculate performance fees properly,
all the transactions done by all investors have to be recorded and
utilized when we perform the corresponding computations to levy the
fees. To give an example, an investor could be holding 10,000 tokens
of the Alpha fund in his wallet, but he could have done several buy
and sell transactions - at different prices - to arrive at this overall
position. That is, a position for an investor is comprised of multiple
transactions. Taking into account the transactions from different
investors implies a large amount of transaction level data. 

The management fee is relatively simpler to calculate and can be done
periodically on the total value locked in the fund when rebalance
events are done - as discussed in Algorithm (\ref{alg:Sequence-Steps-algorithm}).
To calculate performance fees as part of the algorithm - Algorithm
(\ref{alg:Sequence-Steps-algorithm}) - a significant software infrastructure
becomes necessary, which then has to be triggered in one of the corresponding
steps. 

In traditional hedge funds, accountants look at all investor positions
- also the transactions as necessary - every quarter - or other such
infrequent intervals - and calculate how much performance fees are
due. Most hedge funds allow new investments - or redemptions - only
at certain times. This fund entry and exit is usually not very often,
perhaps at quarterly, or longer durations. Hence the tasks of the
fund accountants are relatively straightforward and they do performance
fee calculations easily using specialized software, or even using
excel based tools. 

When the operations of a hedge fund become transparent and accessible
to a wide audience - with the connectivity and visibility provided
by blockchain technology - there will be many transactions happening.
Also to provide the liquidity of mutual funds we have to allow investors
to deposit or withdraw funds at regular intervals, perhaps once a
day. This leads to a lot of investor transactions that become necessary
to calculate performance fees. When investors have the flexibility
to arrive and leave frequently, a similar time frame has to be maintained
for calculating performance fees. Hence we need frequent performance
fee calculations to be happening and having to do these computations
in a blockchain environment compounds the issue.

We next discuss the nuances of calculating the performance fees and
the blockchain constraints that make this a challenging problem.

\subsection{\label{subsec:Blockchain-Bottlenecks-and}Blockchain Bottlenecks
and Making Smart Contracts Smarter}

There are two broad approaches to blockchain performance fee calculations.
They can be done off-chain or on-chain. Off-chain computations refer
to using software routines or calculations outside the blockchain.
On-chain calculations are done entirely within the blockchain environment.
As we know from the foundations of computing science, if we store
a lot of data in the computer memory we can reduce the data we have
to store external to the computer on databases, or other file based
storage systems. If we do not store data within the computer, we have
to expend resources reading and writing data to external infrastructure.
Also, the more data we have, the more intensive the calculations,
or the computation time of the corresponding algorithms increases
with increasing data size. Hence, we need more sophisticated, or complex,
algorithms when we have to handle large data-sets (Knuth 1973; 1997;
Aho \& Hopcroft 1974; Horowitz \& Sahni 1982; Aho 2012).

The problem with off-chain calculations are two fold. One is that
there is less visibility in the calculations. Investors are not able
to see easily the details regarding the fees they are getting charged.
It is not easy to reflect calculations done outside onto the blockchain
environment. Many issues - including security breaches - can arise
due to the use of off-chain software infrastructure. This lack of
security and decentralization defeats the very fundamental reasons
for using blockchain technology. 

Secondly when performance fee calculations - or other computations
- are done outside blockchain they have to be synchronized with on-chain
infrastructure with secondary computations or some synchronization
mechanism. This part of bringing a simplified state of external transactions
- so that fees can be levied at appropriate times, such as when doing
rebalance events - onto the blockchain is not very simple (Eberhardt
\& Tai 2017; 2018; López-Pimentel et al., 2020; Mühlberger et al.,
2020; Emami et al., 2023). 

On-chain calculations are done in smart contracts - or within the
blockchain infrastructure, which then needs to be validated by blockchain
validators. This can be extremely costly - in terms of gas fees (Tikhomirov
2018; Pierro \& Rocha 2019; Faqir-Rhazoui et al., 2021; Donmez \&
Karaivanov 2022; Laurent et al., 2022) - especially when large amounts
of data need to be retrieved - or stored - and calculations have to
be done on such large data-sets. 

It would be extremely computationally demanding to do transaction
level performance fee calculations on-chain. It is resource intensive
to record wallet information - or the address for all investors and
to capture all the transactions done from all the different addresses
- and to use that information for various calculations including the
levy of performance fees. This resource requirement is in terms of
both computational times and memory requirements especially with the
limitations of most blockchain networks (Zheng et al., 2017; Ismail
\& Materwala 2019; Syed et al., 2019; Fan et al., 2020). 

Despite the issues with both approaches - on-chain and off-chain -
the benefits of blockchain accrue only if we make use of the decentralized
infrastructure for our computation needs. Going the on-chain way also
brings reassurance to investors in terms of making it transparent
to them about how they are paying for the services they receive combined
with the cryptographic security of blockchain. We can do on-chain
computations if we find some ways to capture the essence of large
data-sets needed for the specific purpose at hand. In our case, this
is about finding a way to extract the necessary information from a
lot of transactions towards calculating performance fees. 

Kashyap (2021) has a discussion of certain novel security features
and overall architectural designs that have been developed to perform
off-chain computations - given the current blockchain performance
capabilities - for portfolio optimization and risk management purposes.
Such portfolio management decisions are greatly aided by off-chain
calculation routines. It is important to emphasize that these operations
are part of the intrinsic mechanisms of running a portfolio - some
of which provide the fund competitive advantages and rely on proprietary
trading strategies - with less direct need for constant investor scrutiny.
These intensive costly calculations cannot be done on-chain at this
time without significant simplifications - that could affect how the
fund performs in terms of risk and outperforming the market or other
benchmarks - and having them off-chain also ensures that trading strategies
are not easily replicated by others. Though, moving these on-chain
at a later stage should not be ruled out completely.

\subsection{\label{subsec:High-Water-Markets}High Water Markets During Low Market
Prices}

An alternative to storing all transactions from all investors separately
- transaction level processing - on the blockchain, we can keep one
record for each investor that combines the information from all the
transactions done by that investor. Essentially this greatly reduces
the overall number of records we need to work with. This simplification
- wallet or address level processing - still requires us to iterate
through all investors every time we do performance fee calculations.
We would still need to maintain a HWM for each investor so that each
investor obtains the desired level of protection.

An additional improvement would be to combine transactions across
investors and hence aggregate different investor transactions. Instead
of calculating and maintaining the high water mark at the individual
wallet - address level or the position level, that is an aggregation
of the transaction level data - we present various alternatives below,
which keep track of only one HWM benchmark for all investors. Another
reason for pursuing this approach is because in the DeFi world performing
wallet level levy of fees without an action from the user is not possible.
If we do wallet level calculations of performance fees, we must synchronize
the levy of fees to whenever the owner of the wallet deposits or withdraws
from the investment fund. A wallet - and even a smart contract - is
essentially a unique blockchain address with the corresponding owners
being the ones that have access to the private keys to that address. 

We can view blockchain operations as read or write transactions in
public or private domains (Dinh et al., 2018; Thakkar et al., 2018;
Bhushan \& Sharma 2021; End-note \ref{enu:A-public-private-blockchain}).
We understand this further with respect to a wallet belonging to an
investor on a public blockchain. Read transactions read the state
of the blockchain and can be done at anytime without any cost by anyone.
Reading the contents of a wallet - which is reading the state of the
blockchain associated with that wallet - can be done by anyone. Write
transactions change the state of the blockchain and there is a cost
involved to effect this change. Writing or changing the contents of
a wallet - which is changing the state of the blockchain associated
with that wallet - can be done only by the owner of that wallet by
paying gas fees. 

Most of the actions required below can be performed depending on when
the user enters or exits the fund or for the entire TVL at other times
without having to access the individual wallet, or know the corresponding
information. When the investor performs a deposit or withdraw user
action driven, either cash or tokens are being sent to the fund and
they can be used towards fees. We can simply maintain one high water
mark (HWM) - and the corresponding time when the HWM is reached -
for the entire fund. The main issue with this fund level HWM approach
is that once a HWM is established - and the net asset value (NAV)
is below that for a long time period - the new entrants will not pay
a performance fee till the NAV goes above the HWM (End-notes \ref{EN:Net-Asset-Value};
\ref{enu:High-water-mark}). 

There are two broad categories of solutions that can be used to overcome
the issue of the HWM being above the NAV for long periods of time:
\begin{enumerate}
\item The first solution category clubs together different transactions
- both across different investors and within the transactions from
an investor - and maintains some sort of benchmark price or averaged
(weighted average) high water mark corresponding to the positions
that are aggregated together (Section \ref{subsec:Performance-Fees-Across-Aggregate}). 
\item The second solution category keeps investor transactions - and hence
positions - separately and tracks the high water mark separately for
the transactions - and positions - that are kept separate (Section
\ref{subsec:Performance-Fees-Across-Separate}).
\end{enumerate}

\subsection{\label{subsec:Performance-Fees-Across-Aggregate}Performance Fees
Across Aggregated Positions}

This first solution category is \textbf{recommended} since it is closely
aligned to most existing DeFi protocols that club transactions together
across a wallet or address. Section (\ref{subsec:Performance-Fees-Across-Separate})
has a discussion of alternatives to this section which are not recommended
but are given for completeness to help obtain a deeper understanding
of the concepts involved.

\textbf{\textit{Whenever we rebalance - or perform the periodic sequence
of steps (Section \ref{sec:Sequence-of-Steps}) - we calculate the
performance fees and update the high water mark to the new NAV if
the NAV (Alpha price) is above the high water mark. The performance
fee each time is a liability that is used to adjust the NAV accordingly.}}

There are three further possibilities using this approach where we
club together positions and maintain a HWM for the overall clubbed
position. 
\begin{enumerate}
\item \label{Investor-Club-HWM}In the first case, we club together certain
transactions for some investors - or wallet addresses - based on the
following criteria. The group of investor transactions we merge together
would depend on whether these transactions happened when the current
NAV is below the corresponding HWM. We maintain the fund NAV and also
track one HWM for the entire fund. But we keep track of a weighted
average NAV for the transactions that enter when the NAV is below
the HWM. We also track the amount of tokens that have entered below
the HWM. 
\begin{itemize}
\item The performance fees are calculated based on the NAV and the number
of tokens at each rebalance event discussed in Section (\ref{sec:Sequence-of-Steps}). 
\item There will be a performance fee at the fund level and also at the
level of the transactions that have been combined together. The clubbed
transactions fees depends on the transactions that enter when the
NAV is below the HWM. 
\item There will also be performance fees that apply to the tokens that
are being withdrawn and this fees depends on the weighted average
NAV that is being tracked for the transactions that enter when the
NAV is below the HWM. 
\item The advantage of this method compared to the second scenario in Point
(\ref{Fund-Club-HWM}) - which is discussed next - is that we still
have a reference HWM for the entire fund. Having the HWM provides
a reference point which helps with understanding the performance fees
calculations and also explaining to investors more easily how the
performance fees applies to them. 
\item For ease of understanding, we can view this approach as clubbing together
only transactions for a particular investor - that have entered when
the NAV is below the HWM - and hence maintaining the weighted average
NAV for the transactions corresponding to that investor. We can also
club together several investor transactions and use this method across
all those transactions. This would mean that we have to maintain a
weighted averaged NAV for all the transactions - for the relevant
investors - that enter when the NAV is below the HWM. 
\item The extent of granularity desired would depend on how many variables
- and memory - we wish to utilize. Greater the granularity the easier
it is to see and understand the corresponding computations. The trade-off
would depend on the specific blockchain network being used and the
corresponding gas fees and other system dependencies.
\end{itemize}
\item \label{Fund-Club-HWM}In the second case, we club together transactions
- and hence the positions - across all investors or wallet addresses.
We maintain the fund NAV and only a weighted average NAV that applies
to all the tokens that are in the fund right now. 
\begin{itemize}
\item The performance fees are calculated based on the NAV and the number
of tokens at each rebalance event discussed in Section (\ref{sec:Sequence-of-Steps}). 
\item The weighted average NAV and the number of tokens are also updated
at each rebalance event. The performance fees are calculated only
at the fund level and there are no performance fees for the tokens
that are being withdrawn since the entire performance fees liability
is levied at the fund level. 
\item The advantage of this method compared to the first case in Point (\ref{Investor-Club-HWM})
is that we have less calculations to perform and less variables to
track.
\end{itemize}
\item \label{enu:Fund-Put-Back}In the third case, we increase the accuracy
of performance fees collection so that it is identical to collecting
fees based on individual transactions. With the first and second cases
- Points (\ref{Investor-Club-HWM}; \ref{Fund-Club-HWM}) - there
is a possibility of missing out on performance fees when a certain
scenario occurs. 
\begin{itemize}
\item This happens when when the NAV is below the HWM and transactions -
or positions - are made when the weighted average NAV is above the
NAV. Then if the NAV goes up - but still remains lower than the weighted
average NAV - and some withdraw transactions happen they will not
get charged a performance fee. These positions, that are getting redeemed,
have enjoyed positive performance since the price they entered is
lower than the price they are exiting, but they are paying a performance
fee.
\item The performance fees are calculated based on the NAV and the number
of tokens at each rebalance event discussed in Section (\ref{sec:Sequence-of-Steps}). 
\item We handle this scenario by keeping track of two additional variables
that track the NAV movement as follows. We would to know the previous
NAV at each time period and we would need to note down the weighted
average NAV when the NAV dips below the weighted average NAV and then
subsequently raises further - but still remains below the weighted
average NAV. 
\item Solutions given in Points (\ref{Investor-Club-HWM}; \ref{Fund-Club-HWM})
are weighted average simplifications and would suffice when a lot
of transactions are happening and markets do not enter into prolonger
downtrends with partial uptrends. But with the use of two additional
variables - and what we have outlined above - this additional scenario
can also be covered.
\end{itemize}
\end{enumerate}
We discuss all the cases in Points (\ref{Investor-Club-HWM}; \ref{Fund-Club-HWM};
\ref{enu:Fund-Put-Back}) in greater detail below in Sections (\ref{subsec:Investor-Level-Clubbed};
\ref{subsec:Fund-Level-Clubbed}; \ref{subsec:Complete-Solution-including}).

\subsubsection{\label{subsec:Investor-Level-Clubbed}Investor Level Clubbed Positions}

\label{enu:Investor-Clubbed-Positions}We discuss the first scenario
- where we aggregate transactions that have happened when the NAV
is below the HWM - for one investor or wallet address. The case for
adding all investors is a simple extension of this and would be about
performing aggregations and weighted average calculations across transactions
from all investors. 
\begin{itemize}
\item Figure (\ref{fig:Performance-Fees-Illustration:WABHWM}) in Section
(\ref{sec:Numerical-Results}) shows numerical examples related to
this method. This illustration makes it clear why this approach requires
more computations and storage, but can help with easier understanding,
compared to the simpler approach discussed in Point (\ref{Fund-Club-HWM})
in Section (\ref{subsec:Performance-Fees-Across-Aggregate}) and the
discussion in Section (\ref{subsec:Fund-Level-Clubbed}).
\begin{itemize}
\item Since transactions are clubbed together, the issues that need to be
addressed carefully are related to when an investor adds or removes
money at any time and the NAV information to be used to calculate
the corresponding performance fees. 
\item We need additional indicators that track how much of any new money
invested - since the last HWM was established - is above its point
of entry in terms of fund price and hence liable for a performance
fee. Also, we need to know how much of the invested money entered
before the HWM was established to ensure that it does not get charged
multiple performance fees. 
\item We keep track of the cumulative sum of money that gets invested by
any investor when the NAV is below the HWM. We also calculate and
maintain a weighted average NAV per investor (or wallet) since the
last HWM was established.
\item \textbf{We provide the formula (Equation: \ref{eq:HWM-LIABILITY})
for the HWM Liability, $HWML_{jt}$, for investor $j$ at time $t$.}
\begin{align}
HWML_{jt} & =\max\left[0,\left(NAV_{t}-HWM_{t}\right)\right]*PFP_{t}*\left(AMOUNTTOTAL_{jt}-AMOUNTBHWM_{jt}\right)\\
 & +\max\left[0,\left(NAV_{t}-NAVWAVGLHWM_{jt}\right)\right]*PFP_{t}*\left(AMOUNTBHWM_{jt}\right)\label{eq:HWM-LIABILITY}
\end{align}
\end{itemize}
Here, $NAV_{t}$ is the net asset value of the fund at time $t$.
This is also to be understood as the price of the fund at time $t$.
All investors invest in or exit the fund at one or more NAV prices.
Section (\ref{sec:Sequence-of-Steps}) provides details on the steps
regarding how the NAV is updated periodically.

$HWM_{t}$ represents the HWM at time $t$. It is to be understood
that this is the last HWM established before time $t$. $HWMLE_{t}$
is the time of the last HWM event until time $t$. This is the time
when the last HWM was established until time $t$. That is the time,
$t=HWMLE_{t}$, when the last HWM was established or the time when
the HWM was last updated.

$PFP_{t}$ is the performance fee percentage at time $t$. 

$AMOUNTTOTAL_{jt}$ is the total amount in USD corresponding to investor
$j$ invested at time $t$. It is to be understood that each investor
can have multiple transactions - adding or withdrawing money - that
result in this net positive position at time $t$.
\begin{align}
AMOUNTTOTAL_{jt} & =\sum_{l=0}^{NRBE_{t}}\left[INVEST_{jT\left(l\right)}+WITHDRAW_{jT\left(l\right)}\right]
\end{align}
$INVEST_{jt}$ is the amount in USD corresponding to investor $j$
being invested at time $t$. $INVEST_{jt}\geq0$. $WITHDRAW_{jt}$
is the amount in USD corresponding to investor $j$ being withdrawn
at time $t$. $WITHDRAW_{jt}\leq0$. If an investor $j$ does not
participate in a rebalancing event at time $t$, $INVEST_{jt}=WITHDRAW_{jt}=0$. 

At any rebalancing event, only one of the two will be non-zero if
netting across deposits and redemptions are done before the rebalancing.
We retain two variables since it is useful for some of the below formulations.
For simplicity, we are assuming that the investor will only invest
once between rebalance events, since the rebalancing frequency is
quite high.

$T\left(n\right)$ is the time when the $n^{th}$ rebalancing was
done. $n$ is a natural number that includes 0 and is less than or
equal to the number of rebalancing events until a particular time,
$t$. We set $T\left(0\right)=0$ and $T\left(-1\right)=0$.

$NRBE_{t}$ is the total number of rebalancing events until a particular
time, $t$. That is, $0\leq n\leq NRBE_{t}$. Based on our convention,
that time starts $\left(t=0\right)$ when the fund is launched, $NRBE_{t}=0$
until the next rebalancing event happens, which becomes the first
rebalancing event and $NRBE_{t}=1$ after that is completed. \textbf{$NRBE_{t}$
increases by one only after a rebalancing event is completed.} For
example, $NAV_{T\left(NRBE_{t}\right)}$ represents the NAV at the
last rebalancing event until time $t$. 

Note that, $AMOUNTTOTAL_{jt}$ is equivalently calculated using the
below formulae,
\begin{align}
AMOUNTTOTAL_{jt} & =AMOUNTTOTAL_{jT\left(NRBE_{t}-1\right)}\\
 & +\left[INVEST_{jT\left(NRBE_{t}\right)}+WITHDRAW_{jT\left(NRBE_{t}\right)}\right]
\end{align}

$AMOUNTBHWM_{jt}$ is the amount in USD at time $t$ corresponding
to investor $j$ invested after the last HWM, $HWM_{t}$, was established
such that the corresponding subscription prices were below the HWM,
$HWM_{t}$. Note that if the NAV goes up, a new HWM will be established
and the new NAV will become the subscription price. 
\begin{align}
AMOUNTBHWM_{jt} & =\sum_{l=LASTHWMRBE_{t}}^{NRBE_{t}}\left[INVEST_{jT\left(l\right)}\right]\\
 & +\sum_{l=LASTHWMRBE_{t}}^{NRBE_{t}}\left[WITHDRAW_{jT\left(l\right)}\right]
\end{align}
$LASTHWMRBE_{t}$ is the last rebalancing event when the last HWM,
$HWM_{t}$, was established until a particular time, $t$. 

\begin{doublespace}
Note that, $AMOUNTBHWM_{jt}$ is equivalently calculated using the
below formulae,
\begin{align}
AMOUNTBHWM_{jt} & =\sum_{l=0}^{NRBE_{t}}\left[INVEST_{jT\left(l\right)}\right]\left\{ \mathbf{1}(t\geq HWMLE_{t})\right\} \wedge\left\{ \mathbf{1}(NAV_{T\left(l\right)}\leq HWM_{t})\right\} \\
 & +\sum_{l=0}^{NRBE_{t}}\left[WITHDRAW_{jT\left(l\right)}\right]\left\{ \mathbf{1}(t\geq HWMLE_{t})\right\} \wedge\left\{ \mathbf{1}(NAV_{T\left(l\right)}\leq HWM_{t})\right\} \label{eq:AMOUNT-BELOW-HWM-ALT}
\end{align}
\begin{align}
AMOUNTBHWM_{jt} & =AMOUNTBHWM_{jT\left(NRBE_{t}-1\right)}\\
 & +\left[INVEST_{jT\left(NRBE_{t}\right)}+WITHDRAW_{jT\left(NRBE_{t}\right)}\right]\label{eq:AMOUNT-BELOW-HWM}
\end{align}

\end{doublespace}

$\wedge$ represents the ``and'' criteria. $\left\{ \mathbf{1}(A)\right\} $
is the indicator function, which gives $1$ if condition $A$ is TRUE
or $0$ otherwise. 
\begin{equation}
{\displaystyle \mathbf{1}(A):={\begin{cases}
1~ & {\text{ if }}~A\text{ is \textbf{TRUE}}~,\\
0~ & {\text{ if }}~A\text{ is \textbf{FALSE}}~.
\end{cases}}}
\end{equation}
$NAVWAVGLHWM_{jt}$ represents the weighted averaged NAV for investor
$j$ from the time the last HWM, $HWM_{t}$, was established until
a particular time, $t$ based on the amounts invested or withdrawn
by this investor. The formula to calculate this is,
\begin{align}
NAVWAVGLHWM_{jt} & =\left\{ \frac{\sum_{l=LASTHWMRBE_{t}}^{NRBE_{t}}\left[INVEST_{jT\left(l\right)}\right]\left[NAV_{T\left(l\right)}\right]}{\sum_{l=LASTHWMRBE_{t}}^{NRBE_{t}}\left[INVEST_{jT\left(l\right)}\right]}\right\} \label{eq:WAVG-NAV-HWM-ALT}
\end{align}
Note that, $NAVWAVGLHWM_{jt}$ is equivalently calculated using the
below formula,
\begin{align}
NAVWAVGLHWM_{jt} & =\left\{ \frac{\left[NAVWAVGLHWM_{jT\left(NRBE_{t}-1\right)}\right]\sum_{l=LASTHWMRBE_{t}}^{\left(NRBE_{t}-1\right)}\left[INVEST_{jT\left(l\right)}\right]}{\sum_{l=LASTHWMRBE_{t}}^{NRBE_{t}}\left[INVEST_{jT\left(l\right)}\right]}\right\} \\
 & +\left\{ \frac{\left[INVEST_{jT\left(NRBE_{t}\right)}\right]\left[NAV_{T\left(NRBE_{t}\right)}\right]}{\sum_{l=LASTHWMRBE_{t}}^{NRBE_{t}}\left[INVEST_{jT\left(l\right)}\right]}\right\} \label{eq:WVAG-NAV-HWM}
\end{align}
After any rebalancing event is completed, the following updates need
to be done,
\begin{equation}
{\displaystyle HWM_{t}:={\begin{cases}
NAV_{T\left(NBRE_{t}\right)}~ & {\text{ if }}~NAV_{T\left(NBRE_{t}\right)}>HWM_{T\left(NBRE_{t}-1\right)}~,\\
HWM_{T\left(NBRE_{t}-1\right)}~ & {\text{Otherwise}}.
\end{cases}}}
\end{equation}
\begin{equation}
{\displaystyle LASTHWMRBE_{t}:={\begin{cases}
NBRE_{t}~ & {\text{ if }}~NAV_{T\left(NBRE_{t}\right)}>HWM_{T\left(NBRE_{t}-1\right)}~,\\
LASTHWMRBE_{T\left(NBRE_{t}-1\right)}~ & {\text{Otherwise}}.
\end{cases}}}
\end{equation}
\begin{equation}
{\displaystyle AMOUNTBHWM_{jt}:={\begin{cases}
0~ & {\text{ if }}~NAV_{T\left(NBRE_{t}\right)}>HWM_{T\left(NBRE_{t}-1\right)}~,\\
\text{Equation (\ref{eq:AMOUNT-BELOW-HWM-ALT}) Above}~ & {\text{Otherwise}}.
\end{cases}}}
\end{equation}
\begin{equation}
{\displaystyle NAVWAVGLHWM_{jt}:={\begin{cases}
0~ & {\text{ if }}~NAV_{T\left(NBRE_{t}\right)}>HWM_{T\left(NBRE_{t}-1\right)}~,\\
\text{Equation (\ref{eq:WAVG-NAV-HWM-ALT}) Above}~ & {\text{Otherwise}}.
\end{cases}}}
\end{equation}

\end{itemize}

\subsubsection{\label{subsec:Fund-Level-Clubbed}Fund Level Clubbed Positions}

\label{enu:Fund-Clubbed-Positions}The solution in Point (\ref{enu:Investor-Clubbed-Positions})
in Section (\ref{subsec:Investor-Level-Clubbed}) can be significantly
simplified further - without storing the HWM per investor - by simply
updating the weight average NAV price for all the new money that enters
the fund during the current time period. 
\begin{itemize}
\item The performance fee is then simply calculated at the end of every
time period by first updating the NAV - or Fund price - and then checking
if it is above the weighted average NAV that is being maintained.
If the NAV is above the weighted average price, then weighted average
price is set to the NAV that was just calculated. 
\begin{itemize}
\item Storing the HWM provides a narrative which can be easily understood
and explained to investors, but in a DeFi environment, it requires
many additional calculations and gas fees. A simple weighted average
calculation renders the same logic that provides protection to investors
similar to the HWM. 
\item Figure (\ref{fig:Performance-Fees-Illustration:WAFL}) in Section
(\ref{sec:Numerical-Results}) illustrates how this simplified approach
works compared to the approach discussed in Point (\ref{Investor-Club-HWM})
in Section (\ref{subsec:Performance-Fees-Across-Aggregate}) and the
discussion in Section (\ref{enu:Investor-Clubbed-Positions}) with
the corresponding  illustration in Figure (\ref{fig:Performance-Fees-Illustration:WABHWM}).
\end{itemize}
\end{itemize}

\subsubsection{\label{subsec:Complete-Solution-including}Complete Solution Including
the Scenario when NAV Falls and Rises Partially}

The solution outlined here is the most accurate, complete and recommended
solution. The techniques in Points (\ref{Investor-Club-HWM}; \ref{Fund-Club-HWM})
and Sections (\ref{subsec:Investor-Level-Clubbed}; \ref{subsec:Fund-Level-Clubbed})
are simpler to implement and work satisfactorily under most scenarios
except one situation which is considered here in greater detail. 
\begin{itemize}
\item When the NAV falls and rises up later - but stays below the weighted
average NAV - and redemptions happen performance fees will not be
assessed  as outlined in the intuitive explanations provided under
Point (\ref{enu:Fund-Put-Back}). 
\item When the NAV falls and rises up, the deposit transactions that have
happened between the fall and the rise have to incur a performance
fee. But if the NAV stays below the weighted average NAV, and redemptions
happen, they are do not get charged a performance fee. 
\item Figures (\ref{fig:Performance-Fees-Illustration:WABHWM-Fall-Rise-One};
\ref{fig:Performance-Fees-Illustration:WAFL-Fall-Rise-Two}) in Section
(\ref{sec:Numerical-Results}) show how performance fees will not
get charged when the NAV falls - to stay below the weighed NAV - and
rises later but stays below the weighted NAV. 
\item Figures (\ref{fig:Performance-Fees-Illustration:WAFL-Plough-Back-One};
\ref{fig:Performance-Fees-Illustration:WAFL-Plough-Back-Two}) in
Section (\ref{sec:Numerical-Results}) show how the solution presented
in this section handle the scenario shown in Figures (\ref{fig:Performance-Fees-Illustration:WABHWM-Fall-Rise-One};
\ref{fig:Performance-Fees-Illustration:WAFL-Fall-Rise-Two}) relevant
to Points (\ref{Investor-Club-HWM}; \ref{Fund-Club-HWM}) in Sections
(\ref{subsec:Investor-Level-Clubbed}; \ref{subsec:Fund-Level-Clubbed})
regarding how performance fees will not be levied when the NAV falls
- to stay below the weighed NAV - and rises later but stays below
the weighted NAV. 
\item Here, we add the extra conditions to handle the rise and fall scenario
to the Solution outlined in Point (\ref{Investor-Club-HWM}). The
same extension can also easily be applied to the approach in Point
(\ref{Fund-Club-HWM}). 
\item The essential idea to handle this situation is to bring down the weighted
NAV tracker to the level of the NAV after it has risen subsequent
to falling before the rise. When the weighted NAV is lowered the performance
fees corresponding to the lowering is returned back to the fund. Since
we create new tokens when charging performance fees, we have to burn
back tokens when returning the performance fees as discussed in Section
(\ref{sec:Sequence-of-Steps}). 
\item We also levy a performance fee - based on the weighted average and
the new NAV - on the transactions that are getting redeemed.
\item We need to keep track of the weighted average NAV when the NAV is
below the weighted average NAV and the NAV is higher than the NAV
at the previous time period. Using the notation similar to Section
(\ref{subsec:Investor-Level-Clubbed}), we get the following additional
variables we need to monitor.
\begin{equation}
{\displaystyle NAVWAVGPB_{t}:={\begin{cases}
NAV_{T\left(NBRE_{t}\right)}~ & {\text{ if }}\left\{ ~NAV_{T\left(NBRE_{t}\right)}>NAV_{T\left(NBRE_{t}-1\right)}~\right\} \\
 & \wedge\left\{ ~NAVWAVG_{T\left(NBRE_{t}\right)}>NAV_{T\left(NBRE_{t}\right)}~\right\} ,\\
0~ & {\text{Otherwise}}.
\end{cases}}}
\end{equation}
$NAVWAVGPB_{t}$ is the weighted average NAV that keeps track of how
much we need to plough back into the fund at time $t$.

$NAV_{T\left(NBRE_{t}\right)}$ is the NAV at time $T\left(NBRE_{t}\right)$
or the the NAV set at the rebalancing event $NBRE_{t}$.

$NAV_{T\left(NBRE_{t}-1\right)}$ is the NAV at time $T\left(NBRE_{t}-1\right)$
or the the NAV set at the rebalancing event one before $NBRE_{t}$,
that is at the rebalancing event $NBRE_{t}-1$.

$NAVWAVG_{T\left(NBRE_{t}\right)}$ is the weighted average NAV at
time time $T\left(NBRE_{t}\right)$ or the the NAV set at the rebalancing
event $NBRE_{t}$.
\item Note that the weighted average NAV is updated to be $NAVWAVGPB_{t}$
when it is non-zero and is otherwise updated similar to the logic
used in Section (\ref{subsec:Investor-Level-Clubbed}).
\begin{equation}
{\displaystyle NAVWAVG_{T\left(NBRE_{t}\right)}:={\begin{cases}
NAVWAVGPB_{t}~ & {\text{ if }}\left\{ ~NAVWAVGPB_{t}>0~\right\} \\
\text{Similar to Equation (\ref{eq:WAVG-NAV-HWM-ALT}) Above}~ & {\text{Otherwise}}.
\end{cases}}}
\end{equation}
\item The transactions that are getting redeemed are levied a performance
fee according to the formula, 
\begin{align}
RDMPTHWML_{jt} & :={\begin{cases}
\max\left[0,\left(NAV_{T\left(NBRE_{t}\right)}-NAVWAVG_{T\left(NBRE_{t}\right)}\right)\right]\\
*PFP_{t}*\left(AMOUNTRDMPT_{jt}\right)~ & {\text{ if }}\left\{ ~NAVWAVGPB_{t}=0~\right\} \\
\max\left[0,\left(NAV_{T\left(NBRE_{t}\right)}-NAVWAVG_{T\left(NBRE_{t}-1\right)}\right)\right]\\
*PFP_{t}*\left(AMOUNTRDMPT_{jt}\right)~ & {\text{Otherwise}}.
\end{cases}}
\end{align}

$RDMPTHWML_{jt}$ is the performance fees charged on the amount being
redeemed $AMOUNTRDMPT_{jt}$ at time $t$. 
\item The performance fees to be ploughed back - or returned to the fund
- is based on the difference between the weighted average NAV to be
ploughed back and the weighted average NAV at the previous rebalancing
event. It is given according to the formula, 
\begin{align}
PBHWML_{jt} & :={\begin{cases}
0~ & {\text{ if }}\left\{ ~NAVWAVGPB_{t}=0~\right\} \\
\max\left[0,\left(NAVWAVG_{T\left(NBRE_{t}-1\right)}-NAVWAVGPB_{t}\right)\right]\\
*PFP_{t}*\left(AMOUNTTOTAL_{jt}-AMOUNTRDMPT_{jt}\right)~ & {\text{Otherwise}}.
\end{cases}}
\end{align}

$PBHWML_{jt}$ is the performance fees to be returned back at time
$t$. 

$\left(AMOUNTTOTAL_{jt}-AMOUNTRDMPT_{jt}\right)$ gives the amount
still invested in the fund net of the redemption amount at the time
$t$. 
\item The fees that is returned back to the fund will be re-levied as the
NAV improves or as redemptions happen. Time value of money adjustments
can be done if the amount of fees to be returned back are somewhat
large and rebalancing frequencies are not too often (Ross, Westerfield
\& Jaffe 1999; End-note \ref{enu:The-time-value-money}). These time
value adjustments can also be done when the market stays low such
that the NAV is below the weighted average NAV for a long time period.
\end{itemize}

\subsection{\label{subsec:Performance-Fees-Across-Separate}Performance Fees
Across Separate Positions}

This second approach,\textbf{ which is not recommended, is given for
completeness so that we are aware of different alternatives. }This\textbf{
}requires that transactions be kept separate within a wallet or we
need to know the entry and exit price for the transactions - that
make up a position for the address - when the performance fees are
calculated. Section (\ref{subsec:Performance-Fees-Across-Aggregate})
has a discussion of the recommended alternatives to this section.
All the below calculations if done at the transaction level will work
satisfactorily. Whenever we rebalance we calculate the performance
fee if the net asset value (Alpha price) NAV is above the high water
mark and update the high water mark to the new NAV. The performance
fee each time is a liability that is used to adjust the NAV accordingly.
In this case, the HWM above NAV issue can be solved using three options
below:
\begin{enumerate}
\item \label{enu:Free-Ride-Option:}Free Ride Option: We simply let the
new transactions free ride till the fund moves above the HWM. This
acts as an incentive for new money to flow into the fund since the
newly entered investments do not have to pay performance fees till
the fund moves above the HWM. 
\begin{itemize}
\item Also, we can take a performance fee when a transactions is on a free
ride but when that transaction is withdrawn. This is done if the transaction
entered the fund after the time when the previous HWM was established
and the corresponding subscription price (SP) or NAV was below HWM.
When this transaction leaves, if the entry point was below HWM, the
transaction is charged performance fees based on the point of entry
and exit. 
\item Note that a position can have some transactions from before the NAV
has slipped below the HWM. So when redemptions happen some portion
of the transaction may be liable for a performance fee and the rest
might not have any liability. Care has to be taken to charge fees
only for the liable amounts when investors have several transactions.
\item We can also raise the deposit fee for new money if the slump has continued
for a long time. This mitigates the loss of performance fees to a
certain extent.
\end{itemize}
\item \label{enu:New-Money-Liability:}New Money Liability: When new money
enters below the HWM, their liability corresponding to their point
of entry and the HWM is moved to a liability account - and reinvested
into the fund from that account. This is because the new entrant will
not pay any performance fees till the HWM is reestablished. 
\begin{itemize}
\item When a new HWM is established the liabilities are cleared accordingly,
that is we collect the liability outstanding. The liabilities can
also be collected periodically. 
\item Again care has to be taken to ensure that performance fees are only
levied for portions of a position that are made of transactions that
enter when the NAV is below the HWM. 
\item If the investor leaves before a new HWM is reestablished, they pay
corresponding performance fees based on their point of entry and exit.
When this position exits, if the corresponding entry point was below
HWM, they pay performance fees based on their point of entry and exit.
We can also raise the deposit fee for new money if the slump has continued
for a long time. 
\end{itemize}
We provide the formula for the HWM Liability, $HWML_{kt}$, for position
$k$ that is entering at time $t$. It is to be understood that each
position is tracked separately and any investor can have multiple
positions.

\begin{equation}
HWML_{kt}=\max\left(0,HWM_{t}-SP_{kt}\right)*PFP_{t}*INVEST_{kt}*HWMLRATIO_{t}
\end{equation}
Here, $HWM_{t}$ represents the HWM at time $t$. It is to be understood
that this is the last HWM established before or during time $t$. 

$SP_{kt}$ is the NAV that applies to position $k$ when it is subscribing
to (investing in) the fund at time $t$. 

$PFP_{t}$ is the performance fee percentage at time $t$. 

$INVEST_{kt}$ is the amount in USD corresponding to position $k$
being invested at time $t$. It is to be understood that the investor
can have several investments done at multiple time periods. But this
formula applies only to the investment being done at time $t$.

$HWMLRATIO_{t}$ is the HWM Liability ratio percentage at time $t$.
This indicates that we wish to only collect part of the liability.
When this is set to zero, we are providing the free ride option or
charging performance fee upon exit. Note that the performance fee
on exit applies only to those discussed in Option (\ref{enu:Free-Ride-Option:})
above. 

Note that if $HWMLRATIO_{t}=0$ we should collect the performance
fee when the position exits. Care should be taken that either the
HWM liability is charged or the fee is collected upon withdrawal.
So when $HWMLRATIO_{t}=0$ the performance fee is to be charged upon
exit and if $0<HWMLRATIO_{t}\leq1$ the performance fee on exit should
be turned off. There are at-least two sub cases to be taken care to
ensure there is no free riding or overcharging. 
\begin{itemize}
\item When $HWMLRATIO_{t}$ is changed to zero at time $t$, the liabilities
should be returned since there will be an exit fee after time $t$. 
\item When $HWMLRATIO_{t}$ is changed to more than zero at time $t$, this
is difficult to handle and some entrants will get free rides. These
are the entrants who came after the HWM was established, but did not
pay the HWM liability when they entered and now they will not pay
the exit fee since $0<HWMLRATIO_{t}\leq1$. But we can ignore this
case for now for ease of implementation.
\item Note that we do not need a case when NAV is above HWM and positions
enter at the high NAV. This is because we charge performance fees
only when we rebalance and new money enters only when we rebalance.
Hence, new money will enter at the NAV at that time, which will become
the new HWM. This is the reason also why frequent rebalancing, along
with performance fee calculations, is recommended in addition to other
risk management benefits.
\item There could be other cases, so it is important to pay careful attention
to understand the above paths.
\end{itemize}
\item \label{enu:Lower-HWM-and}Lower HWM and Compensate Older Positions:
If the slump continues for a long time and we do not wish to utilize
Option (\ref{enu:New-Money-Liability:}). The third option is that
we lower the HWM and compensate the older positions - and corresponding
transactions across investors - based on their point of entry and
the HWM for which they have paid the fees. The way to do this would
be to calculate for each transaction its point of entry and the corresponding
performance fees. Then some compensation scheme for them has to be
worked out. This could be an airdrop, or, any scheme with a reward
plan that extends over several months (Harrigan et al., 2018; Li et
al., 2024; Allen et al., 2023; End-note \ref{enu:An-airdrop-is}).
This can be very expensive both on the treasury and also in terms
of gas fees, manual intervention needed perhaps etc.
\end{enumerate}
\begin{itemize}
\item A combination of Options (\ref{enu:Free-Ride-Option:}) and (\ref{enu:New-Money-Liability:})
is the most recommended when transaction level information is used.
Option (\ref{enu:Lower-HWM-and}) is the least recommended.
\end{itemize}
The above are meant to be helpful guidelines. Many cases and error
conditions need to be handled appropriately during implementation.
Alternate time conventions and counters are possible and can be accommodated
accordingly. There might even be issues with the counters and timing.
Constructing detailed examples for different cases can help identify
and eliminate any issues. Such issues arise due to the limitations
of not actually testing scenarios using a software system (Beizer
1984; Sneed \& Merey 1985; Livson 1988; Kajihara, Amamiya \& Saya
1993; Bertram et al., 2010; Khanjani \& Sulaiman 2011).

\section{\label{sec:Fund-Flow-Flow}Periodic Blockchain Fund Management Algorithm:
Fund Flow Flow Chart}
\begin{figure}[H]
\includegraphics[width=18cm]{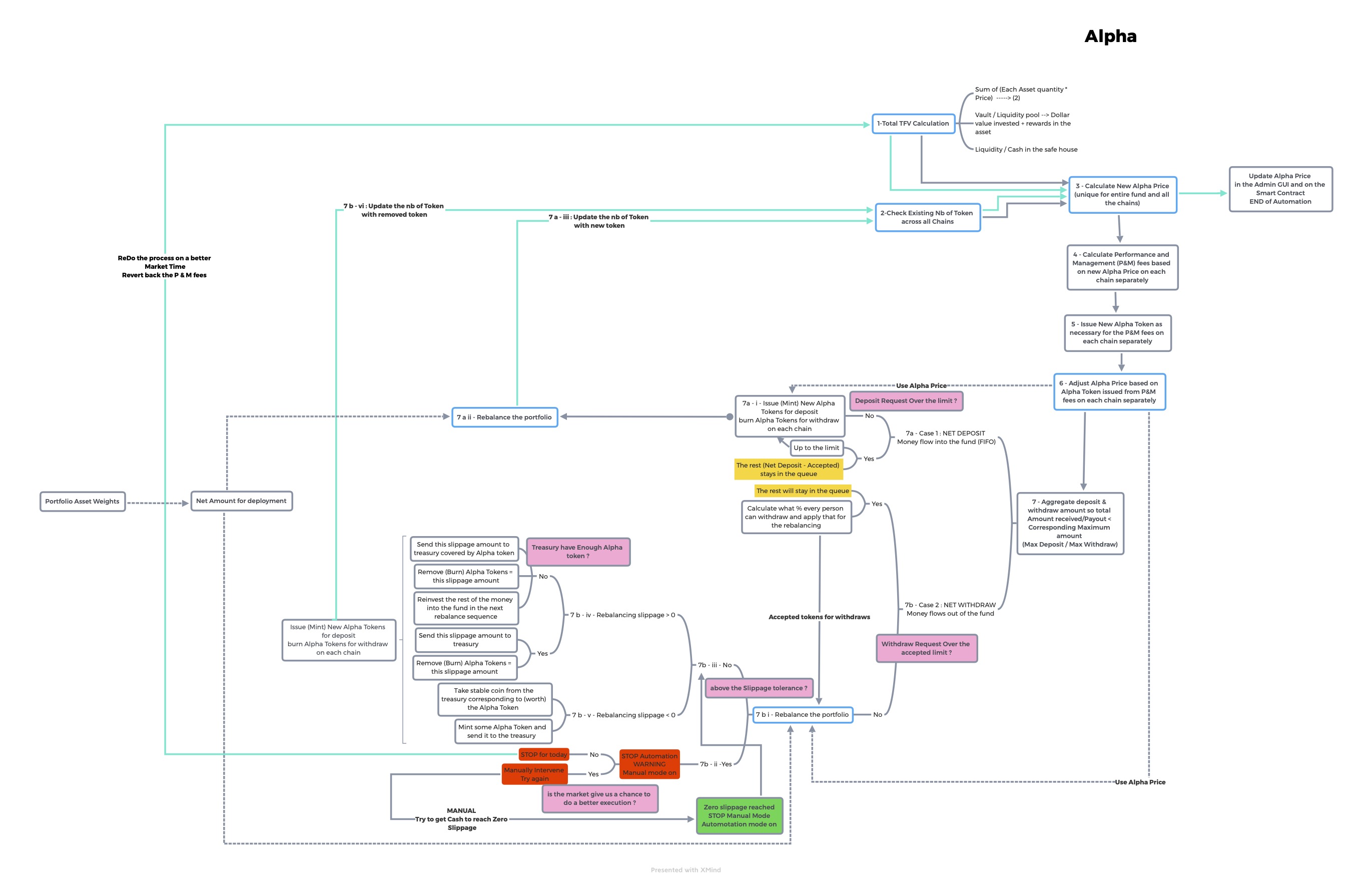}

\caption{Fund Flow Flow Chart: Sequences of Steps for Periodic Fund Management\label{fig:Sequences-of-Steps}}
\end{figure}

\begin{itemize}
\item The flow chart in Figure (\ref{fig:Sequences-of-Steps}) is a visual
illustration corresponding to all the steps mentioned in Algorithm
(\ref{alg:Sequence-Steps-algorithm}) in Section (\ref{sec:Sequence-of-Steps}). 
\item Figures (\ref{fig:Sequences-of-Steps-Input-Variables}; \ref{fig:Sequences-of-Steps-System-Variables};
\ref{fig:Sequences-of-Steps-Deposit-Withdraw-Accept}) in Section
(\ref{sec:Numerical-Results}) give several scenarios pertaining to
the steps and calculations given in Section (\ref{sec:Sequence-of-Steps})
and illustrated with the flow chart here in Figure (\ref{fig:Sequences-of-Steps}).
The figures illustrate different variables corresponding to inputs
and calculated variables, which should help in monitoring how the
system is performing.
\end{itemize}

\section{\label{sec:Numerical-Results}Numerical Results}

\begin{figure}[H]
\includegraphics[width=18cm]{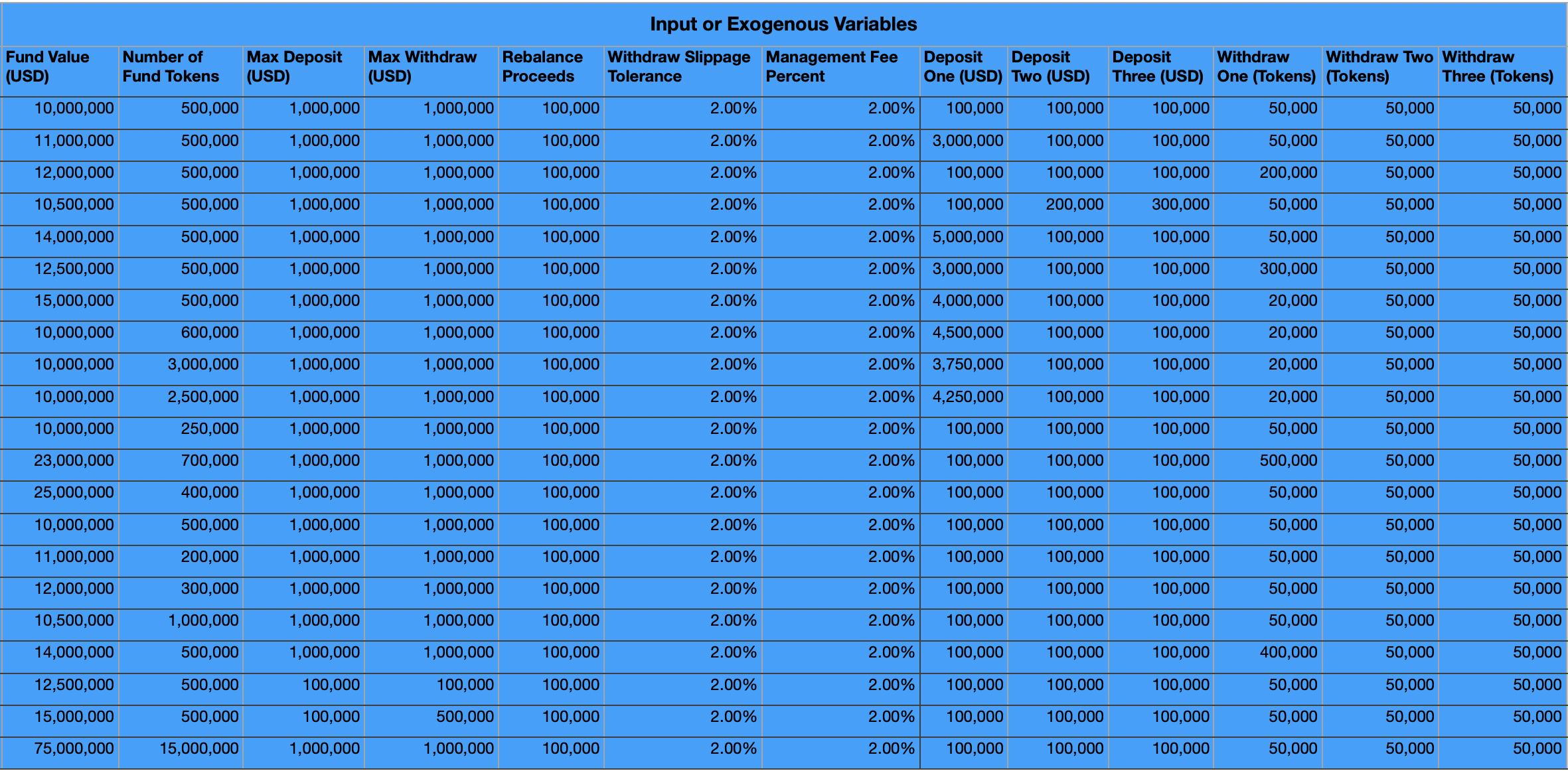}\caption{Sequences of Steps for Periodic Fund Management: Input and Exogenous
Variables\label{fig:Sequences-of-Steps-Input-Variables}}
\end{figure}

\begin{figure}[H]
\includegraphics[width=18cm]{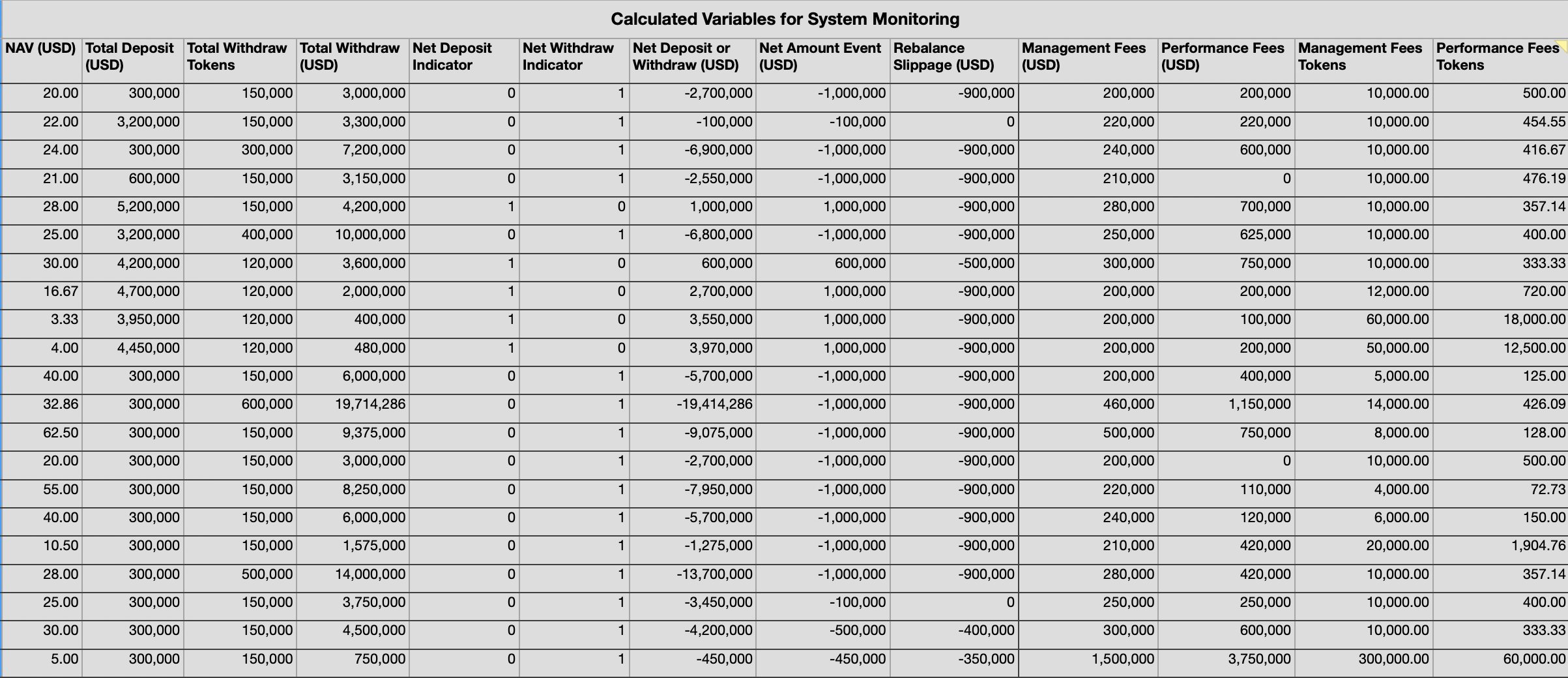}\caption{Sequences of Steps for Periodic Fund Management: Calculated Variables
for System Monitoring\label{fig:Sequences-of-Steps-System-Variables}}
\end{figure}

\begin{figure}[H]
\includegraphics[width=18cm]{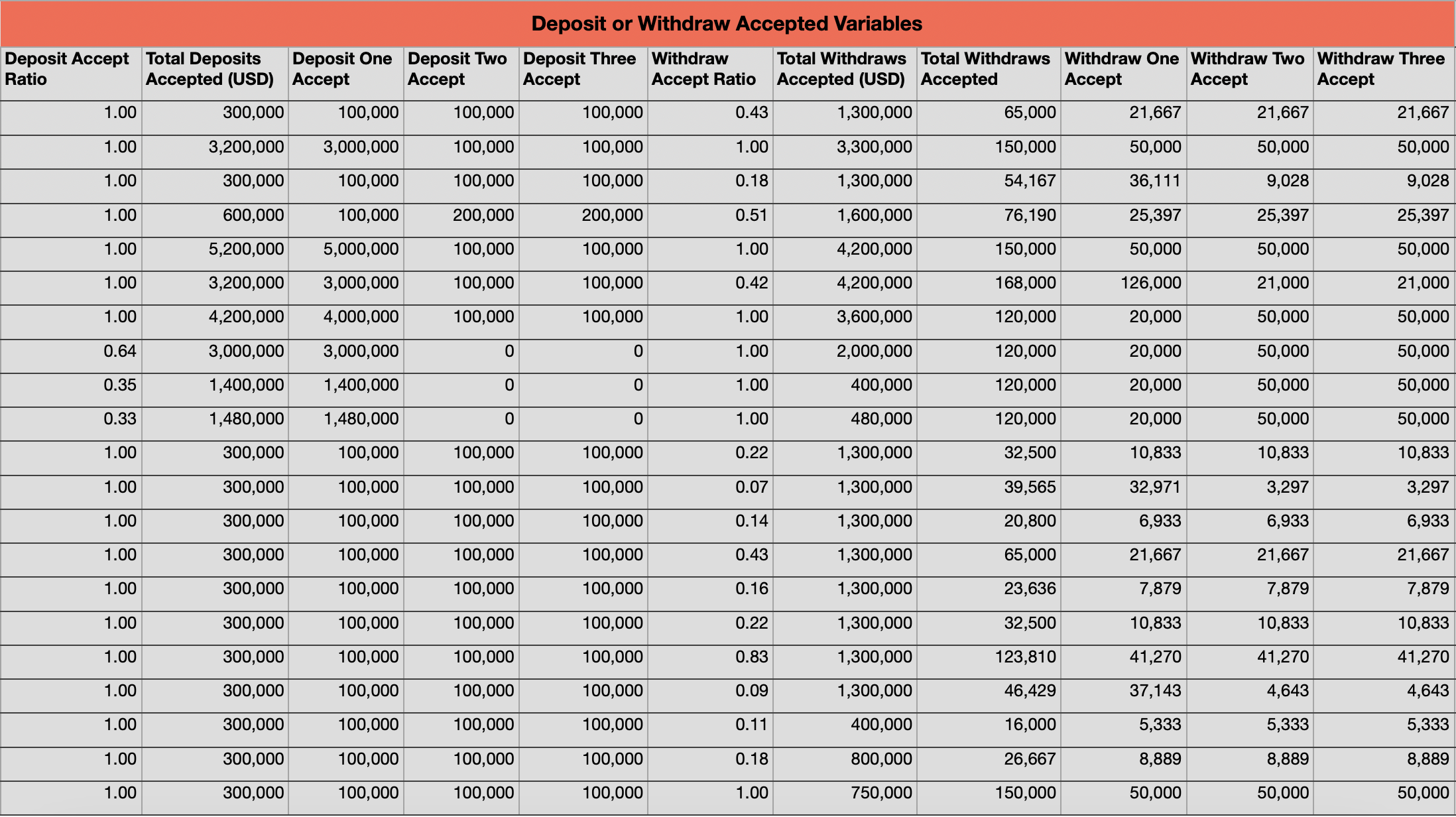}\caption{Sequences of Steps for Periodic Fund Management: Deposit and Withdraw
Accept Variables\label{fig:Sequences-of-Steps-Deposit-Withdraw-Accept}}
\end{figure}

Each of the tables in this section are referenced in the main body
of the article. Below, we provide supplementary descriptions for each
table. 
\begin{itemize}
\item The Table in Figure (\ref{fig:Sequences-of-Steps-Input-Variables})
shows numerical examples related to the sequence of steps discussed
in Section (\ref{sec:Sequence-of-Steps}). 
\item The variables in Figure (\ref{fig:Sequences-of-Steps-Input-Variables})
are input variables which can be changed - or exogenous variables
such as the fund value, which cannot be changed but simulated to understand
its impact on the system and its corresponding performance - to control
how the system is working. 
\item The output calculated variables are given in Figures (\ref{fig:Sequences-of-Steps-System-Variables};
\ref{fig:Sequences-of-Steps-Deposit-Withdraw-Accept}) which can be
observed to gauge system performance and the input variables can be
tweaked to obtain the desired outcomes. Figure (\ref{fig:Sequences-of-Steps-System-Variables})
are general system variables and Figure (\ref{fig:Sequences-of-Steps-Deposit-Withdraw-Accept})
has variable related to how deposit and withdraw requests are being
handled.
\item It should be understood that the values in Figure (\ref{fig:Sequences-of-Steps-Input-Variables})
are just before the rebalance event is about to begin. Many of these
variables can be changed once the system is implemented and we have
given several different scenarios corresponding to different rows
by changing some of these variables in each row. The variables that
cannot be changed are: fund value - which depends on market prices;
number of tokens - which depends on deposits, redemptions, fees and
how the system evolves; rebalance proceeds - market and trading dependent;
and deposits and withdraw requests - which depend on client interactions
and their preferences.
\item The columns in Figure (\ref{fig:Sequences-of-Steps-Input-Variables})
represent the following information respectively: 
\begin{enumerate}
\item \textbf{Scenario Number} gives the scenario given in this row, which
corresponds to one rebalance event given in to Algorithm (\ref{alg:Sequence-Steps-algorithm})
in Section (\ref{sec:Sequence-of-Steps}). 
\item \textbf{Fund Value (USD)} gives the total value of this fund at this
rebalance event. 
\item \textbf{Number of Fund Tokens} gives the number of tokens issued by
the fund at this rebalance event.
\item \textbf{Max Deposit (USD)} is the maximum deposit amount in USD we
accept for investment into the fund at this rebalance event.
\item \textbf{Max Withdraw (USD)} is the maximum amount in USD that can
be sold from the fund to redeem investors at this rebalance event.
\item \textbf{Rebalance Proceeds} are the proceeds obtained from trading
- denominated in USD - at this rebalance event. 
\item \textbf{Withdraw Slippage Tolerance} is a tolerance level we can set
in percentage that shows how much withdraw related trading slippage
we are willing to accept at this rebalance event. 
\item \textbf{Management Fee Percent} is the percent value of management
fees that will be charged annually on the fund value under management.
\item \textbf{Deposit One (USD)} is the deposit request made from investor
one in USD. We assume there are three deposit and withdraw requests
for simplicity. Also if proper netting is implemented the deposit
and withdraw investors will be different.
\item \textbf{Deposit Two (USD) }is the deposit request made from investor
two in USD. 
\item \textbf{Deposit Three (USD) }is the deposit request made from investor
three in USD. 
\item \textbf{Withdraw One (Tokens) }is the withdraw request made from investor
one in number of tokens. 
\item \textbf{Withdraw Two (Tokens) }is the withdraw request made from investor
two in number of tokens. 
\item \textbf{Withdraw Three (Tokens)} is the withdraw request made from
investor three in number of tokens. 
\end{enumerate}
\end{itemize}

\begin{itemize}
\item The Table in Figure (\ref{fig:Sequences-of-Steps-System-Variables})
shows numerical examples related to the sequence of steps discussed
in Section (\ref{sec:Sequence-of-Steps}). 
\item The variables in Figure (\ref{fig:Sequences-of-Steps-Input-Variables})
are input variables which can be changed - or exogenous variables
such as the fund value, which cannot be changed but simulated to its
impact on the system and its corresponding performance - to control
how the system is working. 
\item The output calculated variables are given in Figures (\ref{fig:Sequences-of-Steps-System-Variables};
\ref{fig:Sequences-of-Steps-Deposit-Withdraw-Accept}) which can be
observed to gauge system performance and the input variables can be
tweaked to obtain the desired outcomes. Figure (\ref{fig:Sequences-of-Steps-System-Variables})
are general system variables and Figure (\ref{fig:Sequences-of-Steps-Deposit-Withdraw-Accept})
has variable related to how deposit and withdraw requests are being
handled.
\item It should be understood that the values in Figure (\ref{fig:Sequences-of-Steps-System-Variables})
are just after the rebalance event has started and after we have performed
the corresponding calculations.
\item The columns in Figure (\ref{fig:Sequences-of-Steps-System-Variables})
represent the following information respectively: 
\begin{enumerate}
\item \textbf{NAV (USD) }is the fund price or Net Asset Value at this rebalance
event. It is calculated as the total fund value divided by the number
of tokens.
\item \textbf{Total Deposit (USD) }is the total deposit requests - in USD
- received for investment into the fund at this rebalance event.
\item \textbf{Total Withdraw Tokens }is the total withdraw requests - in
number of tokens - received from investors requesting to take money
out of the fund at this rebalance event.
\item \textbf{Total Withdraw (USD) }is the total withdraw requests - in
USD - received from investors requesting to take money out of the
fund at this rebalance event.
\item \textbf{Net Deposit Indicator }is an indicator which is 1 - 0 otherwise
- if we are doing a net deposit or investment into the fund.
\item \textbf{Net Withdraw Indicator }is an indicator which is 1 - 0 otherwise
- if we are doing a net withdraw or outflow from the fund.
\item \textbf{Net Deposit or Withdraw (USD) }is the value of the total amount
in USD either being invested or withdrawn from the fund at this rebalance
event.
\item \textbf{Net Amount Event (USD) }is the value of the total amount in
USD either being invested or withdrawn from the fund after considering
the maximum values we can invest or take out from the fund during
this rebalance event.
\item \textbf{Rebalance Slippage (USD) }is the difference between the rebalance
proceeds and the net amount event.
\item \textbf{Management Fees (USD) }is the management fees collected in
USD at this rebalance event. It is given as the fund value multiplied
by the management fee percent and applied over the duration since
the last rebalance event.
\item \textbf{Performance Fees (USD) }is the performance fees collected
in USD at this rebalance event. Detailed illustrations and several
scenarios for this calculation are given in later Figures. The values
shown here - for simplicity - are obtained by applying a random percentage
value to the total fund value.
\item \textbf{Management Fees Tokens }is the management fees expressed in
tokens. This is added to the total number of fund tokens.
\item \textbf{Performance Fees Tokens }is the performance fees expressed
in tokens. This is added to the total number of fund tokens.
\end{enumerate}
\end{itemize}

\begin{itemize}
\item The Table in Figure (\ref{fig:Sequences-of-Steps-Deposit-Withdraw-Accept})
shows numerical examples related to the sequence of steps discussed
in Section (\ref{sec:Sequence-of-Steps}). 
\item The variables in Figure (\ref{fig:Sequences-of-Steps-Input-Variables})
are input variables which can be changed - or exogenous variables
such as the fund value, which cannot be changed but simulated to its
impact on the system and its corresponding performance - to control
how the system is working. 
\item The output calculated variables are given in Figures (\ref{fig:Sequences-of-Steps-System-Variables};
\ref{fig:Sequences-of-Steps-Deposit-Withdraw-Accept}) which can be
observed to gauge system performance and the input variables can be
tweaked to obtain the desired outcomes. Figure (\ref{fig:Sequences-of-Steps-System-Variables})
are general system variables and Figure (\ref{fig:Sequences-of-Steps-Deposit-Withdraw-Accept})
has variable related to how deposit and withdraw requests are being
handled.
\item It should be understood that the values in Figure (\ref{fig:Sequences-of-Steps-Deposit-Withdraw-Accept})
are just after the rebalance event has started and after we have performed
the corresponding calculations.
\item The columns in Figure (\ref{fig:Sequences-of-Steps-Deposit-Withdraw-Accept})
represent the following information respectively: 
\begin{enumerate}
\item \textbf{Deposit Accept Ratio }is the deposit accept ratio which gives
the percentage of the total requested deposit amount accepted for
investment into the fund at this rebalance event.
\item \textbf{Total Deposits Accepted (USD) }is the total deposits accepted
- in USD - for investment into the fund at this rebalance event.
\item \textbf{Deposit One Accept }is the amount of deposits accepted - in
USD - from investor one into the fund at this rebalance event. 
\item \textbf{Deposit Two Accept }is the amount of deposits accepted - in
USD - from investor one into the fund at this rebalance event. 
\item \textbf{Deposit Three Accept }is the amount of deposits accepted -
in USD - from investor one into the fund at this rebalance event. 
\item \textbf{Withdraw Accept Ratio }is the withdraw accept ratio which
gives the percentage of the total requested withdraw quantity accepted
for redemption at this rebalance event.
\item \textbf{Total Withdraws Accepted (USD) }is the total withdraw amount
accepted - in USD - to be withdrawn from the fund and distributed
to investors at this rebalance event. 
\item \textbf{Total Withdraws Accepted} is the total withdraw amount accepted
- in tokens - to be withdrawn from the fund and distributed to investors
at this rebalance event. 
\item \textbf{Withdraw One Accept }is the number of tokens accepted for
withdrawal from investor one and to be redeemed from the fund at this
rebalance event. 
\item \textbf{Withdraw Two Accept }is the number of tokens accepted for
withdrawal from investor two and to be redeemed from the fund at this
rebalance event. 
\item \textbf{Withdraw Three Accept }is the number of tokens accepted for
withdrawal from investor three and to be redeemed from the fund at
this rebalance event. 
\end{enumerate}
\end{itemize}

\begin{figure}[H]
\includegraphics[width=18cm]{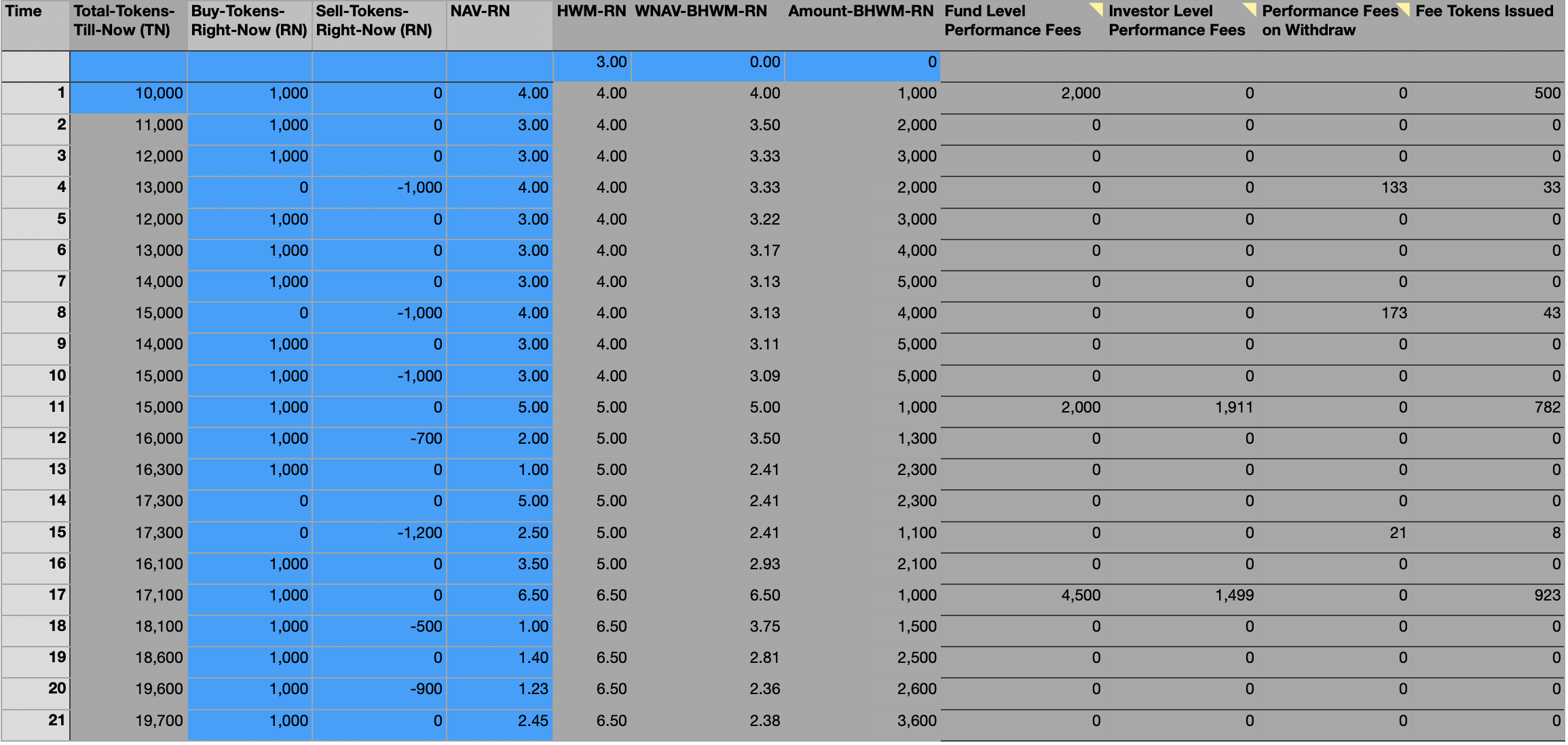}\caption{Performance Fees Illustration: Weighted Average Below High Water Mark\label{fig:Performance-Fees-Illustration:WABHWM}}
\end{figure}

\begin{figure}[H]
\includegraphics[width=18cm]{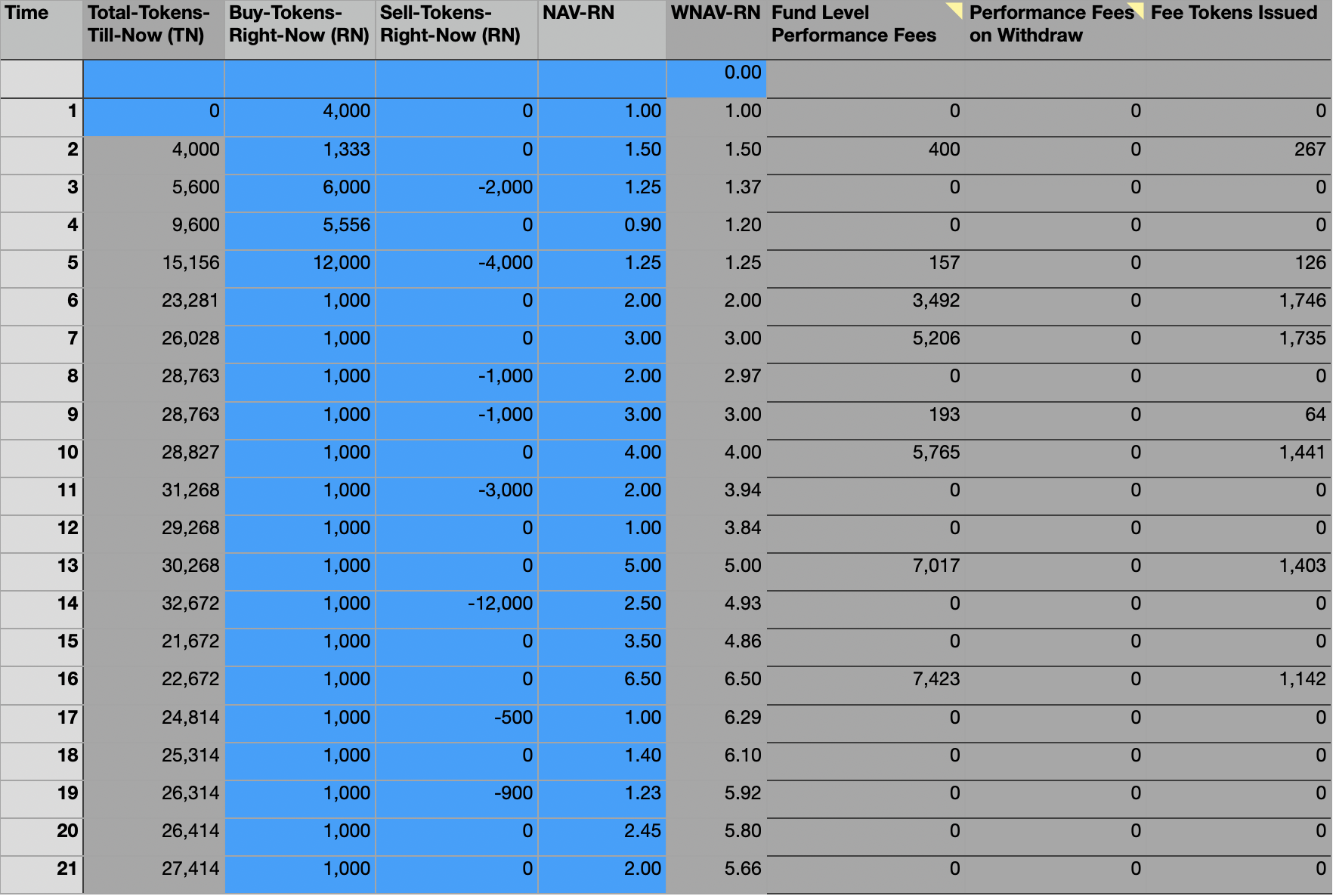}

\caption{Performance Fees Illustration: Weighted Average Fund Level \label{fig:Performance-Fees-Illustration:WAFL}}
\end{figure}

\begin{itemize}
\item The Table in Figure (\ref{fig:Performance-Fees-Illustration:WABHWM})
shows numerical examples related to the method described in Point
(\ref{Investor-Club-HWM}) in Section (\ref{subsec:Performance-Fees-Across-Aggregate})
and the material in Section (\ref{enu:Investor-Clubbed-Positions}).
This illustration makes it clear why this approach requires more computations
and storage, but can help with easier understanding, compared to the
simpler approach from Figure (\ref{fig:Performance-Fees-Illustration:WAFL})
discussed in Point (\ref{Fund-Club-HWM}) in Section (\ref{subsec:Performance-Fees-Across-Aggregate})
and the discussion in Section (\ref{subsec:Fund-Level-Clubbed}).
\item The columns in Figure (\ref{fig:Performance-Fees-Illustration:WABHWM})
represent the following information respectively: 
\begin{enumerate}
\item \textbf{Time} corresponds to each rebalance event corresponding to
Algorithm (\ref{alg:Sequence-Steps-algorithm}) in Section (\ref{sec:Sequence-of-Steps}).
The periodicity could be daily or even intraday intervals.
\item \textbf{Total-Tokens-Till-Now (TN)} shows the total number of tokens
in the fund till this point in time or till now.
\item \textbf{Buy-Tokens-Till-Now (RN)} shows the number of tokens, of the
fund, being bought at this point in time or right now.
\item \textbf{Sell-Tokens-Till-Now (RN)} shows the number of tokens, of
the fund, being sold at this point in time or right now.
\item \textbf{NAV-RN} shows the NAV (Net Asset Value) of the fund at this
point in time or right now.
\item \textbf{HWM-RN} shows the HWM (High Water Mark) of the fund at this
point in time or right now.
\item \textbf{WNAV-BHWM-RN} shows the weighted average NAV for the transactions
corresponding to this investor (or group of investors) that have entered
the fund below the HWM at this point in time or right now.
\item \textbf{Amount-BHWM-RN} shows the total quantity across the transactions
corresponding to this investor (or group of investors) that have entered
the fund below the HWM at this point in time or right now.
\item \textbf{Fund Level Performance Fees }shows the performance fees that
applies to all the tokens in the fund at this point in time or right
now.
\item \textbf{Investor Level Performance Fees }shows the performance fees
that applies to the investor tokens in the fund, across the transactions
corresponding to this investor (or group of investors) that have entered
the fund below the HWM, at this point in time or right now.
\item \textbf{Performance Fees on Withdraw }shows the performance fees that
applies to the tokens being withdraw from the fund at this point in
time or right now.
\item \textbf{Fee Tokens Issued }shows the number of tokens issued corresponding
to the total performance fees that has been levied at this point in
time or right now.
\end{enumerate}
\end{itemize}

\begin{itemize}
\item The Table in Figure (\ref{fig:Performance-Fees-Illustration:WAFL})
shows numerical examples related to the method described in Point
(\ref{Fund-Club-HWM}) in Section (\ref{subsec:Performance-Fees-Across-Aggregate})
and the material in Section (\ref{subsec:Fund-Level-Clubbed}). This
figure illustrates how this simplified approach works compared to
the approach discussed in Point (\ref{Investor-Club-HWM}) in Section
(\ref{subsec:Performance-Fees-Across-Aggregate}) and the discussion
in Section (\ref{enu:Investor-Clubbed-Positions}).
\item The columns in Figure (\ref{fig:Performance-Fees-Illustration:WAFL})
represent the following information respectively: 
\begin{enumerate}
\item \textbf{Time} corresponds to each rebalance event corresponding to
Algorithm (\ref{alg:Sequence-Steps-algorithm}) in Section (\ref{sec:Sequence-of-Steps}).
The periodicity could be daily or even intraday intervals.
\item \textbf{Total-Tokens-Till-Now (TN)} shows the total number of tokens
in the fund till this point in time or till now.
\item \textbf{Buy-Tokens-Till-Now (RN)} shows the number of tokens, of the
fund, being bought at this point in time or right now.
\item \textbf{Sell-Tokens-Till-Now (RN)} shows the number of tokens, of
the fund, being sold at this point in time or right now.
\item \textbf{WNAV-RN} shows the weighted average NAV (Net Asset Value)
of the fund at this point in time or right now.
\item \textbf{Fund Level Performance Fees }shows the performance fees that
applies to all the tokens in the fund at this point in time or right
now.
\item \textbf{Performance Fees on Withdraw }shows the performance fees that
applies to the tokens being withdraw from the fund at this point in
time or right now. This is zero for all scenarios in this simplified
approach.
\item \textbf{Fee Tokens Issued }shows the number of tokens issued corresponding
to the total performance fees that has been levied at this point in
time or right now.
\end{enumerate}
\end{itemize}

\begin{figure}[H]
\includegraphics[width=18cm]{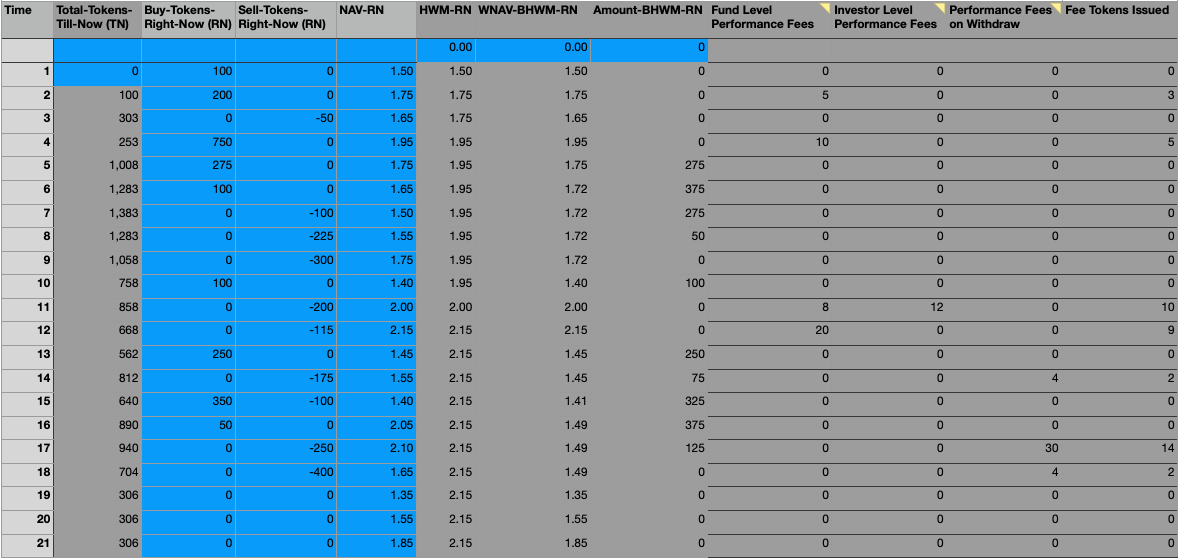}

\caption{Performance Fees Illustration: WA Below HWM Fall-Rise Scenario One\label{fig:Performance-Fees-Illustration:WABHWM-Fall-Rise-One}}
\end{figure}

\begin{figure}[H]
\includegraphics[width=18cm]{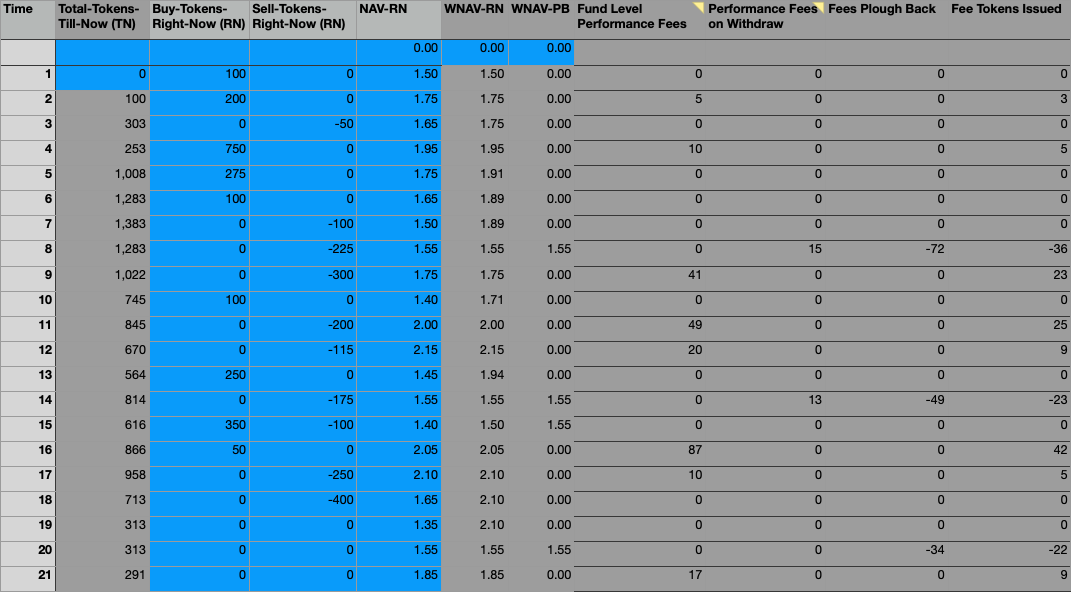}

\caption{Performance Fees Illustration: WAFL Plough Back Fall-Rise Scenario
One\label{fig:Performance-Fees-Illustration:WAFL-Plough-Back-One}}
\end{figure}

\begin{itemize}
\item The Table in Figure (\ref{fig:Performance-Fees-Illustration:WABHWM-Fall-Rise-One})
shows numerical examples related to the rise and fall scenario described
in Point (\ref{enu:Fund-Put-Back}) in Section (\ref{subsec:Performance-Fees-Across-Aggregate}).
The rise and fall scenario is illustrated with respect to the solution
approach outlined in Section (\ref{subsec:Investor-Level-Clubbed}). 
\item The same scenarios are also illustrated in Figure (\ref{fig:Performance-Fees-Illustration:WAFL-Plough-Back-One})
for the solution approach in Section (\ref{subsec:Complete-Solution-including}).
The additional performance fee cash flows in Figure (\ref{fig:Performance-Fees-Illustration:WAFL-Plough-Back-One})
shows that the criteria discussed in Section (\ref{subsec:Investor-Level-Clubbed})
handle the rise and fall situation seamlessly.
\item The columns in Figure (\ref{fig:Performance-Fees-Illustration:WAFL})
represent the same information as the columns in Figure (\ref{fig:Performance-Fees-Illustration:WABHWM})
explained earlier.
\end{itemize}

\begin{figure}[H]
\includegraphics[width=18cm]{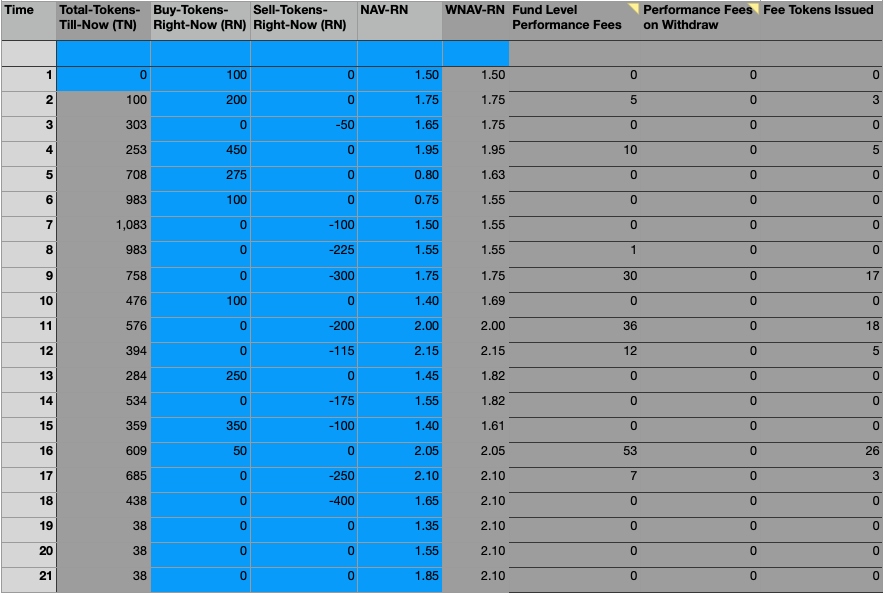}

\caption{Performance Fees Illustration: WA Fund Level Fall-Rise Scenario Two\label{fig:Performance-Fees-Illustration:WAFL-Fall-Rise-Two}}
\end{figure}

\begin{figure}[H]
\includegraphics[width=18cm]{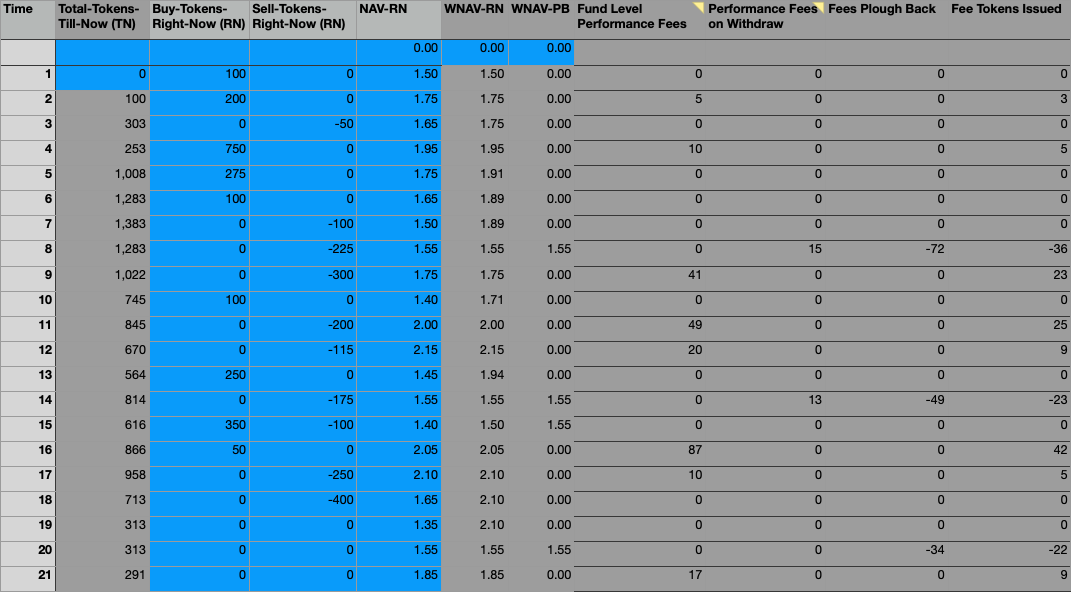}

\caption{Performance Fees Illustration: WAFL Plough Back Fall-Rise Scenario
One\label{fig:Performance-Fees-Illustration:WAFL-Plough-Back-Two}}
\end{figure}

\begin{itemize}
\item The Table in Figure (\ref{fig:Performance-Fees-Illustration:WAFL-Fall-Rise-Two})
shows numerical examples related to the rise and fall scenario described
in Point (\ref{enu:Fund-Put-Back}) in Section (\ref{subsec:Performance-Fees-Across-Aggregate}).
The rise and fall scenario is illustrated with respect to the solution
approach outlined in Section (\ref{subsec:Fund-Level-Clubbed}). 
\item The same scenarios are also illustrated in Figure (\ref{fig:Performance-Fees-Illustration:WAFL-Plough-Back-Two})
for the solution approach in Section (\ref{subsec:Complete-Solution-including}).
The additional performance fee cash flows in Figure (\ref{fig:Performance-Fees-Illustration:WAFL-Plough-Back-Two})
shows that the criteria discussed in Section (\ref{subsec:Investor-Level-Clubbed})
handle the rise and fall situation seamlessly.
\item The columns in Figure (\ref{fig:Performance-Fees-Illustration:WAFL-Fall-Rise-Two})
represent the same information as the columns in Figure (\ref{fig:Performance-Fees-Illustration:WAFL})
explained earlier.
\end{itemize}

\begin{itemize}
\item The Table in Figure (\ref{fig:Performance-Fees-Illustration:WAFL-Plough-Back-One})
shows numerical examples related to the rise and fall scenario described
in Point (\ref{enu:Fund-Put-Back}) in Section (\ref{subsec:Performance-Fees-Across-Aggregate}).
The additional columns - and variables - are added to the solution
approach in Section (\ref{enu:Fund-Clubbed-Positions}). 
\item The same scenarios are also illustrated in Figure (\ref{fig:Performance-Fees-Illustration:WABHWM-Fall-Rise-One})
for the solution approach in Section (\ref{subsec:Investor-Level-Clubbed}).
The additional performance fee cash flows in Figure (\ref{fig:Performance-Fees-Illustration:WAFL-Plough-Back-One})
show that the criteria discussed in Section (\ref{subsec:Investor-Level-Clubbed})
handle the rise and fall situation seamlessly.
\item The columns in Figure (\ref{fig:Performance-Fees-Illustration:WAFL-Plough-Back-One})
represent the same information as the columns in Figure (\ref{fig:Performance-Fees-Illustration:WAFL})
explained earlier, with two additional columns:
\end{itemize}
\begin{enumerate}
\item \textbf{WNAV-PB }shows the weighted average NAV (Net Asset Value)
to be used to plough back - or return - the performance fees back
into the fund as explained in Section (\ref{subsec:Performance-Fees-Across-Aggregate})
and Point (\ref{enu:Fund-Put-Back}) and with the formulations given
in Section (\ref{subsec:Complete-Solution-including}).
\item \textbf{Fees Plough Back }shows the dollar value of the performance
fees to be ploughed back - or returned - into the fund as explained
in Section (\ref{subsec:Performance-Fees-Across-Aggregate}) and Point
(\ref{enu:Fund-Put-Back}) and with the formulations given in Section
(\ref{subsec:Complete-Solution-including}).
\end{enumerate}

\begin{itemize}
\item The Table in Figure (\ref{fig:Performance-Fees-Illustration:WAFL-Plough-Back-Two})
shows numerical examples related to the rise and fall scenario described
in Point (\ref{enu:Fund-Put-Back}) in Section (\ref{subsec:Performance-Fees-Across-Aggregate}).
The additional columns - and variables - are added to the solution
approach in Section (\ref{enu:Fund-Clubbed-Positions}). 
\item The same scenarios are also illustrated in Figure (\ref{fig:Performance-Fees-Illustration:WAFL-Fall-Rise-Two})
for the solution approach in Section (\ref{subsec:Fund-Level-Clubbed}).
The additional performance fee cash flows in Figure (\ref{fig:Performance-Fees-Illustration:WAFL-Plough-Back-Two})
show that the criteria discussed in Section (\ref{subsec:Investor-Level-Clubbed})
handle the rise and fall situation seamlessly.
\item The columns in Figure (\ref{fig:Performance-Fees-Illustration:WAFL-Plough-Back-Two})
represent the same information as the columns in Figure (\ref{fig:Performance-Fees-Illustration:WAFL-Plough-Back-One})
explained earlier.
\end{itemize}

\section{\label{sec:Areas-for-Further}Areas for Further Research}

As better blockchain networks develop, we will need to see if the
above techniques we have created need modification. It might be possible
to use thousands of transactions for calculations on a blockchain
computing platform. Then the need to do weighted averages will be
mitigated since averages are approximations to some extent and having
granular information will yield more accuracy. 

To emphasize, committing thousands of transactions to the blockchain
record, or into a block, is already possible (Pierro \& Tonelli 2022).
The basics of computing make it clear that the more data we wish to
store and the more computations we need to perform, the associated
costs will increase (Dromey 1982). To perform the calculations we
have discussed, using the averaging techniques we have outlined, requires
being able to access a large number of historical transactions as
well. Providing such a large amount of input data to the decentralized
computer is still an area of active research (Wu et al., 2019; Kurt
Peker et al., 2020; Fan, Niu \& Liu 2022; End-note \ref{enu:Ethereum,-which-was}). 

Any investment fund, whether on blockchain or outside, exists to generate
excess returns for its investors. Several excellent investment strategies
have been utilized in traditional investment funds to obtain higher
returns. To implement similar investment ideas on blockchain would
require considering each strategy as an overlay within a larger fund
(Mulvey, Ural \& Zhang 2007; Mohanty, Mohanty \& Ivanof 2021). As
time goes on, several overlay strategies can be added to the basic
fund so that we can benefit from any potential opportunities that
open up. 

A team of researchers and investment specialists need to continually
scour the blockchain investment landscape to identify ways to generate
profits. Another set of bridges that need to be actively built are
strategic partnerships to ensure that the crypto environment can be
highly inclusive, and connect investing to several real world platforms,
solving many problems that plague humanity along the way. Inclusion
of such assets into the portfolio can be similar to socially responsible
investing in the traditional world (Berry \& Junkus 2013; Junkus \&
Berry 2015; End-note \ref{enu:Socially-responsible-investing}) but
with special attention to how such projects might aid the evolution
of the blockchain realm. These will be ongoing and some focus on these
initiatives will be required once the fundamental techniques discussed
here are tested thoroughly and deployed. 

Any intensive computations needed, to clarify the decision process
and arrive at the decision outcomes, can be done outside the blockchain
world, but the essential fund movements are better suited to happen
on a blockchain environment for security reasons. The interaction
between on-chain and off-chain components is a delicate balance involving
several trade-offs such as blockchain computational cost and not revealing
proprietary investment strategies (Garvey \& Murphy 2005; Pardo 2011;
Nuti et al., 2011). 

We have only considered mutual funds and hedge funds as our motivating
vehicles, but other types of funds have numerous innovations that
can be considered in later iterations. For example, exchange traded
funds (ETFs) are a very small extension to what we have discussed
here. The fund tokens provided by the protocols we have discussed
above, can be listed on decentralized exchanges (Jensen, Wachter \&
Ross 2021; Mohan 2022) and the whole system starts behaving like an
ETF. 

Further overlays can be based on specific allocations to sectors we
see as promising. This would be similar to sector themed sub-indices
or ETFs but within a larger grouping of assets (Healy \& Lo 2009;
Mohanty, Mohanty \& Ivanof 2021). These developments can allow investors
to customize their preferences in a basket or theme. Initially it
will be easier to accept investments made only in stable coins (USDT,
USDC and BUSD: End-note \ref{enu:Stablecoin}). We are developing
mechanisms through which investors can participate in blockchain investment
vehicles by making deposits denominated in a larger set of assets
(Kashyap 2021). The stability of stable coins is itself a topic of
significant concern and hence the inclusion of additional assets for
taking deposits, and making redemptions, would be a welcome pursuit
(Hoang \& Baur 2021; Lyons \& Viswanath-Natraj 2023; End-note \ref{enu:LUNA-UST-Collapse}).

As more sophisticated derivatives start to become available as decentralized
securities, incorporating them could be challenging yet rewarding.
The development of new networks, and derivative providers within networks,
will enable the use of options as a hedging mechanism (Hull 2003).
This will help to protect from market crashes and to reduce the portfolio
volatility. Also, derivative strategies combined with rigorous risk
management can help to gain additional returns (Huberts 2004; Madan
\& Sharaiha 2015). Numerous other areas for improvement, in terms
of portfolio weight calculations, rebalancing, trade execution risk
management and so on, are listed in Kashyap (2021). 

\section{\label{sec:Conclusion}Conclusion}

We have created several novel techniques to bring many mechanisms
that have worked well in the traditional financial wealth management
arena to the blockchain space. We have given detailed algorithmic
steps to help with technical implementation of the methodologies we
have developed. Mutual funds, hedge funds and other traditional investments
have had a significant impact in the lives of many individuals across
the world. Despite their popularity, there are many concerns regarding
their transparency and ease of access for everyone. Blockchain technology
is extremely well suited to mitigate - if not entirely eliminate -
those concerns. Decentralized ledger concepts and the technological
advancements over the last several decades allow us to combine the
best features of both hedge funds and mutual funds. 

We have given detailed mathematical formulations, and technical pointers,
to be able to implement the mechanisms we have created as blockchain
smart contracts. Our approach overcomes numerous blockchain bottlenecks
and takes the power of smart contracts much further. We have shown
how fund prices can be updated regularly like mutual funds and performance
fees can be charged like hedge funds. In addition blockchain investment
funds - as we have described - can operate with investor protection
schemes such as high water marks and measures to offset trading related
slippage costs when redemptions happen.

Equal access to transparent wealth creation opportunities for everyone
are finally around the corner.

\section{\label{sec:End-notes}End-notes }
\begin{enumerate}
\item \label{enu:Open-end-mutual-funds}Open-end mutual funds are purchased
from or sold to the issuer at the net asset value of each share as
of the close of the trading day in which the order was placed, as
long as the order was placed within a specified period before the
close of trading. They can be traded directly with the issuer. \href{https://en.wikipedia.org/wiki/Mutual_fund}{Mutual Fund,  Wikipedia Link}
\item \label{enu:A-hedge-fund}A hedge fund is a pooled investment fund
that trades in relatively liquid assets and is able to make extensive
use of more complex trading, portfolio-construction, and risk management
techniques in an attempt to improve performance, such as short selling,
leverage, and derivatives. \href{https://en.wikipedia.org/wiki/Hedge_fund}{Hedge Fund,  Wikipedia Link}
\item \label{enu:Finance-AUM}In finance, assets under management (AUM),
sometimes called fund under management, measures the total market
value of all the financial assets which an individual or financial
institution—such as a mutual fund, venture capital firm, or depository
institution—or a decentralized network protocol controls, typically
on behalf of a client. \href{https://en.wikipedia.org/wiki/Assets_under_management}{Assets Under Management, Wikipedia Link}
\item \label{enu:TVL}In decentralized finance, Total value locked represents
the number of assets that are currently being staked in a specific
protocol.\href{https://coinmarketcap.com/alexandria/glossary/total-value-locked-tvl}{Total Value Locked, CoinMarketCap Link}
\item \label{enu:Decentralized-finance}Decentralized finance (often stylized
as DeFi) offers financial instruments without relying on intermediaries
such as brokerages, exchanges, or banks by using smart contracts on
a blockchain. \href{https://en.wikipedia.org/wiki/Decentralized_finance}{Decentralized Finance (DeFi), Wikipedia Link}
\item \label{enu:Types-Yield-Enhancement-Services}The following are the
four main types of blockchain decentralized financial products or
services. We can also consider them as the main types of yield enhancement,
or return generation, vehicles available in decentralized finance: 
\begin{enumerate}
\item \label{enu:Single-sided-staking-allows}Single-Sided Staking: This
allows users to earn yield by providing liquidity for one type of
asset, in contrast to liquidity provisioning on AMMs, which requires
a pair of assets. \href{https://docs.saucerswap.finance/features/single-sided-staking}{Single Sided Staking,  SuacerSwap Link}
\begin{enumerate}
\item Bancor is an example of a provider who supports single sided staking.
Bancor natively supports Single-Sided Liquidity Provision of tokens
in a liquidity pool. This is one of the main benefits to liquidity
providers that distinguishes Bancor from other DeFi staking protocols.
Typical AMM liquidity pools require a liquidity provider to provide
two assets. Meaning, if you wish to deposit \textquotedbl TKN1\textquotedbl{}
into a pool, you would be forced to sell 50\% of that token and trade
it for \textquotedbl TKN2\textquotedbl . When providing liquidity,
your deposit is composed of both TKN1 and TKN2 in the pool. Bancor
Single-Side Staking changes this and enables liquidity providers to:
Provide only the token they hold (TKN1 from the example above) Collect
liquidity providers fees in TKN1. \href{https://docs.bancor.network/about-bancor-network/faqs/single-side-liquidity}{Single Sided Staking,  Bancor Link}
\end{enumerate}
\item \label{enu:AMM-Liquidity-Pairs}AMM Liquidity Pairs (AMM LP): A constant-function
market maker (CFMM) is a market maker with the property that that
the amount of any asset held in its inventory is completely described
by a well-defined function of the amounts of the other assets in its
inventory (Hanson 2007). \href{https://en.wikipedia.org/wiki/Constant_function_market_maker}{Constant Function Market Maker,  Wikipedia Link}

This is the most common type of market maker liquidity pool. Other
types of market makers are discussed in Mohan (2022). All of them
can be grouped under the category Automated Market Makers. Hence the
name AMM Liquidity Pairs. A more general discussion of AMMs, without
being restricted only to the blockchain environment, is given in (Slamka,
Skiera \& Spann 2012).
\item \label{enu:LP-Token-Staking:}LP Token Staking: LP staking is a valuable
way to incentivize token holders to provide liquidity. When a token
holder provides liquidity as mentioned earlier in Point (\ref{enu:AMM-Liquidity-Pairs})
they receive LP tokens. LP staking allows the liquidity providers
to stake their LP tokens and receive project tokens tokens as rewards.
This mitigates the risk of impermanent loss and compensates for the
loss. \href{https://defactor.com/liquidity-provider-staking-introduction-guide/}{Liquidity Provider Staking,  DeFactor Link}
\begin{enumerate}
\item Note that this is also a type of single sided staking discussed in
Point (\ref{enu:Single-sided-staking-allows}). The key point to remember
is that the LP Tokens can be considered as receipts for the crypto
assets deposits in an AMM LP Point (\ref{enu:AMM-Liquidity-Pairs}).
These LP Token receipts can be further staked to generate additional
yield.
\end{enumerate}
\item \label{enu:Lending:-Crypto-lending}Lending: Crypto lending is the
process of depositing cryptocurrency that is lent out to borrowers
in return for regular interest payments. Payments are typically made
in the form of the cryptocurrency that is deposited and can be compounded
on a daily, weekly, or monthly basis. \href{https://www.investopedia.com/crypto-lending-5443191}{Crypto Lending,  Investopedia Link};
\href{https://defiprime.com/decentralized-lending}{DeFi Lending,  DeFiPrime Link};
\href{https://crypto.com/price/categories/lending}{Top Lending Coins by Market Capitalization,  Crypto.com Link}.
\begin{enumerate}
\item Crypto lending is very common on decentralized finance projects and
also in centralized exchanges. Centralized cryptocurrency exchanges
are online platforms used to buy and sell cryptocurrencies. They are
the most common means that investors use to buy and sell cryptocurrency
holdings. \href{https://www.investopedia.com/tech/what-are-centralized-cryptocurrency-exchanges/}{Centralized Cryptocurrency Exchanges,  Investopedia Link}
\item Lending is a very active area of research both on blockchain and off
chain (traditional finance) as well (Cai 2018; Zeng et al., 2019;
Bartoletti, Chiang \& Lafuente 2021; Gonzalez 2020; Hassija et al.,
2020; Patel et al. , 2020). 
\item Lending is also a highly profitable business in the traditional financial
world (Kashyap 2022-I). Investment funds, especially hedge funds,
engage in borrowing securities to put on short positions depending
on their investment strategies.Long only investment funds typically
supply securities or lend their assets for a fee.
\item In finance, a long position in a financial instrument means the holder
of the position owns a positive amount of the instrument. \href{https://en.wikipedia.org/wiki/Long_(finance)}{Long Position in Finance,  Wikipedia Link}
\item In finance, being short in an asset means investing in such a way
that the investor will profit if the value of the asset falls. This
is the opposite of a more conventional \textquotedbl long\textquotedbl{}
position, where the investor will profit if the value of the asset
rises. \href{https://en.wikipedia.org/wiki/Short_(finance)}{Short Position in Finance,  Wikipedia Link}
\end{enumerate}
\end{enumerate}
\item \label{EN:Net-Asset-Value}Net Asset Value is the net value of an
investment fund's assets less its liabilities, divided by the number
of shares outstanding. \href{https://www.investopedia.com/terms/n/nav.asp}{NAV,  Investopedia Link}
\item \label{enu:An-index-fund}An index fund (also index tracker) is a
mutual fund or exchange-traded fund (ETF) designed to follow certain
preset rules so that it can replicate the performance (\textquotedbl track\textquotedbl )
of a specified basket of underlying investments. \href{https://en.wikipedia.org/wiki/Index_fund}{Index Fund,  Wikipedia Link}
\begin{enumerate}
\item A mutual fund is an investment fund that pools money from many investors
to purchase securities. Mutual Funds typically pay their regular and
recurring, fund-wide operating expenses out of fund assets, rather
than by imposing separate fees and charges directly on investors.
\href{https://www.investor.gov/introduction-investing/investing-basics/glossary/mutual-fund-fees-and-expenses}{Mutual Fund Fees,   Wikipedia Link}
\item An exchange-traded fund (ETF) is a type of investment fund and exchange-traded
product, i.e. they are traded on stock exchanges. ETFs are similar
in many ways to mutual funds, except that ETFs are bought and sold
from other owners throughout the day on stock exchanges whereas mutual
funds are bought and sold from the issuer based on their price at
day's end. \href{https://en.wikipedia.org/wiki/Exchange-traded_fund}{Exchange Traded Fund,  Wikipedia Link}
\end{enumerate}
\item \label{enu:A-smart-contract}A smart contract is a computer program
or a transaction protocol that is intended to automatically execute,
control or document events and actions according to the terms of a
contract or an agreement. The objectives of smart contracts are the
reduction of need for trusted intermediators, arbitration costs, and
fraud losses, as well as the reduction of malicious and accidental
exceptions. Smart contracts are commonly associated with cryptocurrencies,
and the smart contracts introduced by Ethereum are generally considered
a fundamental building block for decentralized finance (DeFi) and
NFT applications. \href{https://en.wikipedia.org/wiki/Smart_contract}{Smart Contract,  Wikipedia Link}
\item \label{enu:High-water-mark}High-water mark is the highest level of
value reached by an investment account or portfolio. It is often used
as a threshold to determine whether a fund manager can gain a performance
fee. Investors benefit from a high-water mark by avoiding paying performance-based
bonuses for poor performance or for the same performance twice. \href{https://corporatefinanceinstitute.com/resources/wealth-management/high-water-mark/}{High Water Mark (HWM), Corporate Finance Institute Link}
\item \label{enu:In-computing-FIFO}In computing and in systems theory,
first in, first out (the first in is the first out), acronymized as
FIFO, is a method for organizing the manipulation of a data structure
(often, specifically a data buffer) where the oldest (first) entry,
or \textquotedbl head\textquotedbl{} of the queue, is processed first.
Such processing is analogous to servicing people in a queue area on
a first-come, first-served (FCFS) basis, i.e. in the same sequence
in which they arrive at the queue's tail. \href{https://en.wikipedia.org/wiki/FIFO_(computing_and_electronics)}{First in First Out (Computing),  Wikipedia Link}
\item \label{enu:In-financial-markets,}In financial markets, implementation
shortfall is the difference between the decision price and the final
execution price (including commissions, taxes, etc.) for a trade.
This is also known as the \textquotedbl slippage\textquotedbl .
Agency trading is largely concerned with minimizing implementation
shortfall and finding liquidity. \href{https://en.wikipedia.org/wiki/Implementation_shortfall}{Implementation Shortfall,  Wikipedia Link}
\item \label{enu:Stablecoin}A Stable-coin is a type of cryptocurrency where
the value of the digital asset is supposed to be pegged to a reference
asset, which is either fiat money, exchange-traded commodities (such
as precious metals or industrial metals), or another cryptocurrency.
\href{https://en.wikipedia.org/wiki/Stablecoin}{Stable Coin,  Wikipedia Link}
\item \label{enu:A-public-private-blockchain}A public blockchain has absolutely
no access restrictions. Anyone with an Internet connection can send
transactions to it as well as become a validator. \href{https://en.wikipedia.org/wiki/Blockchain\#Types}{Blockchain Types,  Wikipedia Link}
\begin{itemize}
\item A private blockchain is permissioned. One cannot join it unless invited
by the network administrators. Participant and validator access is
restricted.
\end{itemize}
\item \label{enu:The-time-value-money}The time value of money is the widely
accepted conjecture that there is greater benefit to receiving a sum
of money now rather than an identical sum later. It may be seen as
an implication of the later-developed concept of time preference.
\href{https://en.wikipedia.org/wiki/Time_value_of_money}{Time Value of Money,  Wikipedia Link}
\item \label{enu:An-airdrop-is}An airdrop is an unsolicited distribution
of a cryptocurrency token or coin, usually for free, to numerous wallet
addresses. \href{https://en.wikipedia.org/wiki/Airdrop_(cryptocurrency)}{Airdrop (Cryptocurrency),  Wikipedia Link}
\item \label{enu:Ethereum,-which-was}Ethereum, which was conceived in 2013
and launched in 2015 (Wood 2014; Tapscott \& Tapscott 2016; Dannen
2017), provided a remarkable innovation in terms of making blockchain
based systems Turing complete (or theoretically being able to do what
any computer can do: Sipser 2006).
\begin{enumerate}
\item Ethereum is a decentralized, open-source blockchain with smart contract
functionality. Ether (Abbreviation: ETH) is the native cryptocurrency
of the platform. Among cryptocurrencies, ether is second only to bitcoin
in market capitalization. \href{https://en.wikipedia.org/wiki/Ethereum}{Ethereum,  Wikipedia Link}
\item \label{enu:Turing-Complete}In computability theory, a system of data-manipulation
rules (such as a computer's instruction set, a programming language,
or a cellular automaton) is said to be Turing-complete or computationally
universal if it can be used to simulate any Turing machine. \href{https://en.wikipedia.org/wiki/Turing_completeness}{Turing Completeness,  Wikipedia Link}

A Turing machine is a mathematical model of computation describing
an abstract machine that manipulates symbols on a strip of tape according
to a table of rules. Despite the model's simplicity, it is capable
of implementing any computer algorithm. \href{https://en.wikipedia.org/wiki/Turing_machine}{Turing Machine,  Wikipedia Link}
\end{enumerate}
\item \label{enu:Socially-responsible-investing}Socially responsible investing
(SRI), social investment, sustainable socially conscious, \textquotedbl green\textquotedbl{}
or ethical investing, is any investment strategy which seeks to consider
both financial return and social/environmental good to bring about
social change regarded as positive by proponents. \href{https://en.wikipedia.org/wiki/Socially_responsible_investing}{Socially Responsible Investing,  Wikipedia Link}
\begin{enumerate}
\item Environmental, social, and corporate governance (ESG) is a framework
designed to be embedded into an organization's strategy that considers
the needs and ways in which to generate value for all of organizational
stakeholders (such as employees, customers and suppliers and financiers).
\href{https://en.wikipedia.org/wiki/Environmental,_social,_and_corporate_governance}{Environmental,  Social and Corporate Governance,  Wikipedia Link}
\item The areas of concern recognized by the SRI practitioners are sometimes
summarized under the heading of ESG issues: environment, social justice,
and corporate governance.
\end{enumerate}
\item \label{enu:LUNA-UST-Collapse}The recent LUNA / UST episode on the
Terra network, from May 8 to May 13 2022 and beyond, is a demonstration
of the risk of holding concentrated portfolios (Lee et al., 2022;
Briola et al. , 2023).
\end{enumerate}

\section{\label{sec:References}References}
\begin{itemize}
\item Agarwal, V., Boyson, N. M., \& Naik, N. Y. (2009). Hedge funds for
retail investors? An examination of hedged mutual funds. Journal of
Financial and Quantitative Analysis, 44(2), 273-305.
\item Agarwal, V., Daniel, N. D., \& Naik, N. Y. (2009). Role of managerial
incentives and discretion in hedge fund performance. The Journal of
Finance, 64(5), 2221-2256.
\item Aggarwal, R. K., \& Jorion, P. (2012). Is there a cost to transparency?.
Financial Analysts Journal, 68(2), 108-123.
\item Agarwal, V., Vashishtha, R., \& Venkatachalam, M. (2018). Mutual fund
transparency and corporate myopia. The Review of Financial Studies,
31(5), 1966-2003.
\item Ahmad, A. H., Green, C., \& Jiang, F. (2020). Mobile money, financial
inclusion and development: A review with reference to African experience.
Journal of economic surveys, 34(4), 753-792.
\item Aho, A. V., \& Hopcroft, J. E. (1974). The design and analysis of
computer algorithms. Pearson Education India. 
\item Aho, A. V. (2012). Computation and computational thinking. The computer
journal, 55(7), 832-835.
\item Aiken, A. L., Clifford, C. P., \& Ellis, J. A. (2015). Hedge funds
and discretionary liquidity restrictions. Journal of Financial Economics,
116(1), 197-218.
\item Allen, D. W., Berg, C., \& Lane, A. M. (2023). Why airdrop cryptocurrency
tokens?. Journal of Business Research, 163, 113945.
\item Andrew, S. T., \& Herbert, B. (2015). Modern operating systems. Pearson
Education.
\item Anson, M. J. (2002). Hedge Fund Transparency. The Journal of Wealth
Management, 5(2), 79-83.
\item Ante, L., Fiedler, I., \& Strehle, E. (2021). The influence of stablecoin
issuances on cryptocurrency markets. Finance Research Letters, 41,
101867.
\item Aragon, G. O. (2007). Share restrictions and asset pricing: Evidence
from the hedge fund industry. Journal of financial economics, 83(1),
33-58.
\item Aragon, G. O., Hertzel, M., \& Shi, Z. (2013). Why do hedge funds
avoid disclosure? Evidence from confidential 13F filings. Journal
of Financial and Quantitative Analysis, 48(5), 1499-1518.
\item Asness, C. S., Krail, R. J., \& Liew, J. M. (2001). Do hedge funds
hedge?. The journal of portfolio management, 28(1), 6-19.
\item Ballandies, M. C., Dapp, M. M., \& Pournaras, E. (2022). Decrypting
distributed ledger design—taxonomy, classification and blockchain
community evaluation. Cluster computing, 25(3), 1817-1838.
\item Bartoletti, M., Chiang, J. H. Y., \& Lafuente, A. L. (2021). SoK:
lending pools in decentralized finance. In Financial Cryptography
and Data Security. FC 2021 International Workshops: CoDecFin, DeFi,
VOTING, and WTSC, Virtual Event, March 5, 2021, Revised Selected Papers
25 (pp. 553-578). Springer Berlin Heidelberg. 
\item Beizer, B. (1984). Software system testing and quality assurance.
Van Nostrand Reinhold Co..
\item Ben-David, I., Birru, J., \& Rossi, A. (2020). The performance of
hedge fund performance fees (No. w27454). National Bureau of Economic
Research.
\item Berry, T. C., \& Junkus, J. C. (2013). Socially responsible investing:
An investor perspective. Journal of business ethics, 112, 707-720.
\item Bertram, D., Voida, A., Greenberg, S., \& Walker, R. (2010, February).
Communication, collaboration, and bugs: the social nature of issue
tracking in small, collocated teams. In Proceedings of the 2010 ACM
conference on Computer supported cooperative work (pp. 291-300).
\item Bertsimas, D., \& Lo, A. W. (1998). Optimal control of execution costs.
Journal of financial markets, 1(1), 1-50.
\item Bhushan, B., \& Sharma, N. (2021). Transaction privacy preservations
for blockchain technology. In International Conference on Innovative
Computing and Communications: Proceedings of ICICC 2020, Volume 2
(pp. 377-393). Springer Singapore.
\item Briola, A., Vidal-Tomás, D., Wang, Y., \& Aste, T. (2023). Anatomy
of a Stablecoin’s failure: The Terra-Luna case. Finance Research Letters,
51, 103358.
\item Broby, D. (2012). The regulatory evolution of the post credit crisis
fund management industry. Journal of Economic Policy Reform, 15(1),
5-11.
\item Brooks, C., \& Kat, H. M. (2002). The statistical properties of hedge
fund index returns and their implications for investors. The Journal
of Alternative Investments, 5(2), 26-44.
\item Brown, S., \& Pomerantz, S. (2017). Some clarity on mutual fund fees.
U. Pa. J. Bus. L., 20, 767. 
\item Buterin, V. (2014). A next-generation smart contract and decentralized
application platform. white paper, 3(37), 2-1.
\item Cai, C. W. (2018). Disruption of financial intermediation by FinTech:
a review on crowdfunding and blockchain. Accounting \& Finance, 58(4),
965-992. 
\item Chen, D. H., \& Huang, H. L. (2018). Panic, slash, or crash—Do black
swans flap in stock markets?. Physica A: Statistical Mechanics and
its Applications, 492, 1642-1663. 
\item Cherkes, M., Sagi, J., \& Stanton, R. (2008). A liquidity-based theory
of closed-end funds. The Review of Financial Studies, 22(1), 257-297. 
\item Chordia, T. (1996). The structure of mutual fund charges. Journal
of financial Economics, 41(1), 3-39.
\item Cici, G., Gibson, S., \& Moussawi, R. (2010). Mutual fund performance
when parent firms simultaneously manage hedge funds. Journal of Financial
Intermediation, 19(2), 169-187.
\item Ciriello, R. F. (2021). Tokenized index funds: A blockchain-based
concept and a multidisciplinary research framework. International
Journal of Information Management, 61, 102400. 
\item Coates IV, J. C., \& Hubbard, R. G. (2007). Competition in the mutual
fund industry: Evidence and implications for policy. J. Corp. L.,
33, 151.
\item Cremers, M., Ferreira, M. A., Matos, P., \& Starks, L. (2016). Indexing
and active fund management: International evidence. Journal of Financial
Economics, 120(3), 539-560.
\item Cumming, D., \& Dai, N. (2009). Capital flows and hedge fund regulation.
Journal of Empirical Legal Studies, 6(4), 848-873.
\item Cuthbertson, K., Nitzsche, D., \& O'Sullivan, N. (2010). Mutual fund
performance: Measurement and Evidence. Financial Markets, Institutions
\& Instruments, 19(2), 95-187.
\item Dannen, C. (2017). Introducing Ethereum and solidity (Vol. 1, pp.
159-160). Berkeley: Apress.
\item Dellva, W. L., \& Olson, G. T. (1998). The relationship between mutual
fund fees and expenses and their effects on performance. Financial
Review, 33(1), 85-104.
\item Deakin, S. (2015). The evolution of theory and method in law and finance.
The Oxford handbook of financial regulation, 14-15. 
\item De Gregorio, J., \& Guidotti, P. E. (1995). Financial development
and economic growth. World development, 23(3), 433-448.
\item Di Pierro, M. (2017). What is the blockchain?. Computing in Science
\& Engineering, 19(5), 92-95.
\item Dinh, T. T. A., Liu, R., Zhang, M., Chen, G., Ooi, B. C., \& Wang,
J. (2018). Untangling blockchain: A data processing view of blockchain
systems. IEEE transactions on knowledge and data engineering, 30(7),
1366-1385.
\item Donmez, A., \& Karaivanov, A. (2022). Transaction fee economics in
the Ethereum blockchain. Economic Inquiry, 60(1), 265-292.
\item Dromey, R. G. (1982). How to Solve it by Computer. Prentice-Hall,
Inc..
\item Eberhardt, J., \& Tai, S. (2017). On or off the blockchain? Insights
on off-chaining computation and data. In Service-Oriented and Cloud
Computing: 6th IFIP WG 2.14 European Conference, ESOCC 2017, Oslo,
Norway, September 27-29, 2017, Proceedings 6 (pp. 3-15). Springer
International Publishing. 
\item Eberhardt, J., \& Tai, S. (2018, July). Zokrates-scalable privacy-preserving
off-chain computations. In 2018 IEEE International Conference on Internet
of Things (iThings) and IEEE Green Computing and Communications (GreenCom)
and IEEE Cyber, Physical and Social Computing (CPSCom) and IEEE Smart
Data (SmartData) (pp. 1084-1091). IEEE. 
\item El Faqir, Y., Arroyo, J., \& Hassan, S. (2020, August). An overview
of decentralized autonomous organizations on the blockchain. In Proceedings
of the 16th international symposium on open collaboration (pp. 1-8).
\item Eling, M., \& Faust, R. (2010). The performance of hedge funds and
mutual funds in emerging markets. Journal of Banking \& Finance, 34(8),
1993-2009.
\item Eling, M., \& Schuhmacher, F. (2007). Does the choice of performance
measure influence the evaluation of hedge funds?. Journal of Banking
\& Finance, 31(9), 2632-2647.
\item Elton, E. J., Gruber, M. J., \& Blake, C. R. (2003). Incentive fees
and mutual funds. The Journal of Finance, 58(2), 779-804.
\item Emami, A., Keshavarz Kalhori, G., Mirzakhani, S., \& Akhaee, M. A.
(2023). A blockchain-based privacy-preserving anti-collusion data
auction mechanism with an off-chain approach. The Journal of Supercomputing,
1-50.
\item Fan, C., Ghaemi, S., Khazaei, H., \& Musilek, P. (2020). Performance
evaluation of blockchain systems: A systematic survey. IEEE Access,
8, 126927-126950.
\item Fan, X., Niu, B., \& Liu, Z. (2022). Scalable blockchain storage systems:
research progress and models. Computing, 104(6), 1497-1524.
\item Faqir-Rhazoui, Y., Ariza-Garzón, M. J., Arroyo, J., \& Hassan, S.
(2021, May). Effect of the gas price surges on user activity in the
daos of the ethereum blockchain. In Extended Abstracts of the 2021
CHI Conference on Human Factors in Computing Systems (pp. 1-7).
\item Feldman, D., Saxena, K., \& Xu, J. (2020). Is the active fund management
industry concentrated enough?. Journal of Financial Economics, 136(1),
23-43.
\item Fiergbor, D. D. (2018). Blockchain technology in fund management.
In Applications of Computing and Communication Technologies: First
International Conference, ICACCT 2018, Delhi, India, March 9, 2018,
Revised Selected Papers 1 (pp. 310-319). Springer Singapore.
\item Fung, W., \& Hsieh, D. A. (1999). A primer on hedge funds. Journal
of empirical finance, 6(3), 309-331.
\item Garvey, R., \& Murphy, A. (2005). Entry, exit and trading profits:
A look at the trading strategies of a proprietary trading team. Journal
of Empirical Finance, 12(5), 629-649.
\item Giancaspro, M. (2017). Is a ‘smart contract’really a smart idea? Insights
from a legal perspective. Computer law \& security review, 33(6),
825-835.
\item Gil‐Bazo, J., \& Ruiz‐Verdú, P. A. B. L. O. (2009). The relation between
price and performance in the mutual fund industry. The Journal of
Finance, 64(5), 2153-2183.
\item Goetzmann, W. N., Ingersoll Jr, J. E., \& Ross, S. A. (2003). High‐water
marks and hedge fund management contracts. The Journal of Finance,
58(4), 1685-1718.
\item Goetzmann, W. N., \& Rouwenhorst, K. G. (2005). The origins of value:
The financial innovations that created modern capital markets. Oxford
University Press, USA.
\item Golec, J. H. (1996). The effects of mutual fund managers' characteristics
on their portfolio performance, risk and fees. Financial Services
Review, 5(2), 133-147.
\item Goltz, F., \& Schröder, D. (2010). Hedge fund transparency: where
do we stand?. The Journal of Alternative Investments, 12(4), 20-35.
\item Gonzalez, L. (2020). Blockchain, herding and trust in peer-to-peer
lending. Managerial Finance, 46(6), 815-831. 
\item Grassi, L., Lanfranchi, D., Faes, A., \& Renga, F. M. (2022). Do we
still need financial intermediation? The case of decentralized finance–DeFi.
Qualitative Research in Accounting \& Management.
\item Guasoni, P., \& Obłój, J. (2016). The incentives of hedge fund fees
and high‐water marks. Mathematical Finance, 26(2), 269-295.
\item Hanson, R. (2007). Logarithmic market scoring rules for modular combinatorial
information aggregation. The Journal of Prediction Markets, 1(1),
3-15. 
\item Harrigan, M., Shi, L., \& Illum, J. (2018, November). Airdrops and
privacy: a case study in cross-blockchain analysis. In 2018 IEEE International
Conference on Data Mining Workshops (ICDMW) (pp. 63-70). IEEE.
\item Haslem, J. A. (2004). A tool for improved mutual fund transparency.
The Journal of Investing, 13(3), 54-64.
\item Haslem, J. A. (2007). Normative transparency of mutual fund disclosure
and the case of the expense ratio. The Journal of Investing, 16(4),
167-174.
\item Hassija, V., Bansal, G., Chamola, V., Kumar, N., \& Guizani, M. (2020).
Secure lending: Blockchain and prospect theory-based decentralized
credit scoring model. IEEE Transactions on Network Science and Engineering,
7(4), 2566-2575. 
\item Healy, A. D., \& Lo, A. W. (2009). Jumping the gates: Using beta-overlay
strategies to hedge liquidity constraints. Journal of investment management,
3(11).
\item Hedges IV, J. R. (2005). Hedge fund transparency. The European Journal
of Finance, 11(5), 411-417.
\item Hoang, L. T., \& Baur, D. G. (2021). How stable are stablecoins?.
The European Journal of Finance, 1-17.
\item Hong, X. (2014). The dynamics of hedge fund share restrictions. Journal
of Banking \& Finance, 49, 82-99.
\item Horowitz, E., \& Sahni, S. (1982). Fundamentals of data structures.
\item Hu, B., Zhang, Z., Liu, J., Liu, Y., Yin, J., Lu, R., \& Lin, X. (2021).
A comprehensive survey on smart contract construction and execution:
paradigms, tools, and systems. Patterns, 2(2).
\item Huang, Y., Bian, Y., Li, R., Zhao, J. L., \& Shi, P. (2019). Smart
contract security: A software lifecycle perspective. IEEE Access,
7, 150184-150202.
\item Huberts, L. C. (2004). Overlay Speak. The Journal of Investing, 13(3),
22-30.
\item Hull, J. C. (2003). Options futures and other derivatives. Pearson
Education India. 
\item Huynh, T. D., \& Xia, Y. (2023). Panic selling when disaster strikes:
Evidence in the bond and stock markets. Management Science, 69(12),
7448-7467.
\item Ismail, L., \& Materwala, H. (2019). A review of blockchain architecture
and consensus protocols: Use cases, challenges, and solutions. Symmetry,
11(10), 1198.
\item Jensen, J. R., von Wachter, V., \& Ross, O. (2021). An introduction
to decentralized finance (defi). Complex Systems Informatics and Modeling
Quarterly, (26), 46-54.
\item Junkus, J., \& Berry, T. D. (2015). Socially responsible investing:
a review of the critical issues. Managerial Finance, 41(11), 1176-1201.
\item Kajihara, J., Amamiya, G., \& Saya, T. (1993). Learning from bugs
(software quality control). IEEE Software, 10(5), 46-54.
\item Kannengießer, N., Lins, S., Sander, C., Winter, K., Frey, H., \& Sunyaev,
A. (2021). Challenges and common solutions in smart contract development.
IEEE Transactions on Software Engineering, 48(11), 4291-4318.
\item Kashyap, R. (2020). David vs Goliath (You against the Markets), A
dynamic programming approach to separate the impact and timing of
trading costs. Physica A: Statistical Mechanics and its Applications,
545, 122848.
\item Kashyap, R. (2021). A Tale of Two Currencies: Cash and Crypto. Available
at SSRN 4400742.
\item Kashyap, R. (2022-I). Options as Silver Bullets: Valuation of Term
Loans, Inventory Management, Emissions Trading and Insurance Risk
Mitigation using Option Theory. Annals of Operations Research, 315(2),
1175-1215.
\item Kavanagh, D., Lightfoot, G., \& Lilley, S. (2014). Finance past, finance
future: a brief exploration of the evolution of financial practices.
Management \& Organizational History, 9(2), 135-149.
\item Khanjani, A., \& Sulaiman, R. (2011, March). The process of quality
assurance under open source software development. In 2011 IEEE Symposium
on Computers \& Informatics (pp. 548-552). IEEE.
\item Khorana, A., Servaes, H., \& Tufano, P. (2009). Mutual fund fees around
the world. The Review of Financial Studies, 22(3), 1279-1310.
\item Kitamura, T. (2010). A model on the financial panic. Physica A: Statistical
Mechanics and its Applications, 389(13), 2586-2596.
\item Kitzler, S., Victor, F., Saggese, P., \& Haslhofer, B. (2023). Disentangling
decentralized finance (DeFi) compositions. ACM Transactions on the
Web, 17(2), 1-26.
\item Kleinnijenhuis, J., Schultz, F., Oegema, D., \& Van Atteveldt, W.
(2013). Financial news and market panics in the age of high-frequency
sentiment trading algorithms. Journalism, 14(2), 271-291.
\item Knuth, D. E. (1973). The art of computer programming (Vol. 3). Reading,
MA: Addison-Wesley. 
\item Knuth, D. E. (1997). The Art of Computer Programming: Fundamental
Algorithms, volume 1. Addison-Wesley Professional.
\item Kruse, R. L. (1984). Data structures \& program design.
\item Kurt Peker, Y., Rodriguez, X., Ericsson, J., Lee, S. J., \& Perez,
A. J. (2020). A cost analysis of internet of things sensor data storage
on blockchain via smart contracts. Electronics, 9(2), 244.
\item Laurent, A., Brotcorne, L., \& Fortz, B. (2022). Transaction fees
optimization in the Ethereum blockchain. Blockchain: Research and
Applications, 3(3), 100074.
\item Lee, S., Lee, J., \& Lee, Y. (2022). Dissecting the Terra-LUNA crash:
Evidence from the spillover effect and information flow. Finance Research
Letters, 103590.
\item Lenkey, S. L. (2015). The closed-end fund puzzle: Management fees
and private information. Journal of Financial Intermediation, 24(1),
112-129. 
\item Levine, R. (1997). Financial development and economic growth: views
and agenda. Journal of economic literature, 35(2), 688-726. 
\item Levine, R. (2005). Finance and growth: theory and evidence. Handbook
of economic growth, 1, 865-934.
\item Levinthal, D., \& Myatt, J. (1994). Co‐evolution of capabilities and
industry: the evolution of mutual fund processing. Strategic Management
Journal, 15(S1), 45-62. 
\item Li, J., Wan, X., Cheng, H. K., \& Zhao, X. (2024). Operation dumbo
drop: To airdrop or not to airdrop for initial coin offering success?.
Information Systems Research.
\item Liang, B. (1999). On the performance of hedge funds. Financial Analysts
Journal, 55(4), 72-85.
\item Liu, Z., Li, Y., Min, Q., \& Chang, M. (2022). User incentive mechanism
in blockchain-based online community: An empirical study of steemit.
Information \& Management, 59(7), 103596.
\item Livson, B. U. (1988). A practical approach to software quality assurance.
ACM SIGSOFT Software Engineering Notes, 13(3), 45-48. 
\item Lo, A. (2017). Adaptive markets: Financial evolution at the speed
of thought. Princeton University Press.
\item López-Pimentel, J. C., Rojas, O., \& Monroy, R. (2020, November).
Blockchain and off-chain: A solution for audit issues in supply chain
systems. In 2020 IEEE International Conference on Blockchain (Blockchain)
(pp. 126-133). IEEE.
\item Lyons, R. K., \& Viswanath-Natraj, G. (2023). What keeps stablecoins
stable?. Journal of International Money and Finance, 131, 102777.
\item Macrinici, D., Cartofeanu, C., \& Gao, S. (2018). Smart contract applications
within blockchain technology: A systematic mapping study. Telematics
and Informatics, 35(8), 2337-2354.
\item Madan, D. B., \& Sharaiha, Y. M. (2015). Option overlay strategies.
Quantitative Finance, 15(7), 1175-1190.
\item Malkiel, B. G., \& Saha, A. (2005). Hedge funds: Risk and return.
Financial analysts journal, 61(6), 80-88.
\item Malkiel, B. G. (2013). Asset management fees and the growth of finance.
Journal of Economic Perspectives, 27(2), 97-108.
\item Mallaby, S. (2010). More money than god: Hedge funds and the making
of the new elite. A\&C Black.
\item Manda, V. K., Manda, V. K., \& Katneni, V. (2023). Blockchain for
the asset management industry. World Review of Science, Technology
and Sustainable Development, 19(1-2), 170-185.
\item Manda, V. K., \& SS, P. Rao. (2018, November). Blockchain technology
for the mutual fund industry. In National Seminar on Paradigm Shifts
in Commerce and Management (pp. 12-17). 
\item Matz, L., \& Neu, P. (Eds.). (2006). Liquidity risk measurement and
management: A practitioner's guide to global best practices (Vol.
408). John Wiley \& Sons. 
\item Minsky, H. P. (1986). The evolution of financial institutions and
the performance of the economy. Journal of Economic Issues, 20(2),
345-353. 
\item Minsky, H. P. (1990). Schumpeter: finance and evolution. Evolving
technology and market structure: Studies in Schumpeterian economics,
51-73. 
\item Mohan, V. (2022). Automated market makers and decentralized exchanges:
a DeFi primer. Financial Innovation, 8(1), 20.
\item Mohanta, B. K., Panda, S. S., \& Jena, D. (2018, July). An overview
of smart contract and use cases in blockchain technology. In 2018
9th international conference on computing, communication and networking
technologies (ICCCNT) (pp. 1-4). IEEE.
\item Mohanty, S. S., Mohanty, O., \& Ivanof, M. (2021). Alpha enhancement
in global equity markets with ESG overlay on factor-based investment
strategies. Risk Management, 23(3), 213-242.
\item Mühlberger, R., Bachhofner, S., Castelló Ferrer, E., Di Ciccio, C.,
Weber, I., Wöhrer, M., \& Zdun, U. (2020). Foundational oracle patterns:
Connecting blockchain to the off-chain world. In Business Process
Management: Blockchain and Robotic Process Automation Forum: BPM 2020
Blockchain and RPA Forum, Seville, Spain, September 13–18, 2020, Proceedings
18 (pp. 35-51). Springer International Publishing.
\item Mulvey, J. M., Ural, C., \& Zhang, Z. (2007). Improving performance
for long-term investors: wide diversification, leverage, and overlay
strategies. Quantitative Finance, 7(2), 175-187.
\item Nakamoto, S. (2008). Bitcoin: A peer-to-peer electronic cash system.
Decentralized business review, 21260.
\item Negara, E. S., Hidayanto, A. N., Andryani, R., \& Syaputra, R. (2021).
Survey of smart contract framework and its application. Information,
12(7), 257.
\item Nuti, G., Mirghaemi, M., Treleaven, P., \& Yingsaeree, C. (2011).
Algorithmic trading. Computer, 44(11), 61-69.
\item Pardo, R. (2011). The evaluation and optimization of trading strategies.
John Wiley \& Sons.
\item Patel, S. B., Bhattacharya, P., Tanwar, S., \& Kumar, N. (2020). Kirti:
A blockchain-based credit recommender system for financial institutions.
IEEE Transactions on Network Science and Engineering, 8(2), 1044-1054. 
\item Patel, R., Migliavacca, M., \& Oriani, M. E. (2022). Blockchain in
banking and finance: A bibliometric review. Research in International
Business and Finance, 62, 101718. 
\item Penman, S. H. (1970). What Net Asset Value?-{}-An Extension of a Familiar
Debate. The Accounting Review, 45(2), 333-346.
\item Philippon, T. (2015). Has the US finance industry become less efficient?
On the theory and measurement of financial intermediation. American
Economic Review, 105(4), 1408-1438.
\item Pierro, G. A., \& Rocha, H. (2019, May). The influence factors on
ethereum transaction fees. In 2019 IEEE/ACM 2nd International Workshop
on Emerging Trends in Software Engineering for Blockchain (WETSEB)
(pp. 24-31). IEEE.
\item Pierro, G. A., \& Tonelli, R. (2022, March). Can Solana be the solution
to the blockchain scalability problem?. In 2022 IEEE International
Conference on Software Analysis, Evolution and Reengineering (SANER)
(pp. 1219-1226). IEEE.
\item Prat, A. (2005). The wrong kind of transparency. American economic
review, 95(3), 862-877.
\item Rajan, R., \& Zingales, L. (1998). Financial development and growth.
American economic review, 88(3), 559-586. 
\item Ren, Y. S., Ma, C. Q., Chen, X. Q., Lei, Y. T., \& Wang, Y. R. (2023).
Sustainable finance and blockchain: A systematic review and research
agenda. Research in International Business and Finance, 101871. 
\item Renshaw, E. F. (1984). Stock market panics: a test of the efficient
market hypothesis. Financial Analysts Journal, 40(3), 48-51.
\item Ross, S. A., Westerfield, R., \& Jaffe, J. F. (1999). Corporate finance.
Irwin/McGraw-Hill.
\item Saurabh, K., Rani, N., \& Upadhyay, P. (2023). Towards blockchain
led decentralized autonomous organization (DAO) business model innovations.
Benchmarking: An International Journal, 30(2), 475-502.
\item Sayeed, S., Marco-Gisbert, H., \& Caira, T. (2020). Smart contract:
Attacks and protections. IEEE Access, 8, 24416-24427.
\item Sharma, P., Jindal, R., \& Borah, M. D. (2023). A review of smart
contract-based platforms, applications, and challenges. Cluster Computing,
26(1), 395-421.
\item Silber, W. L. (1983). The process of financial innovation. The American
Economic Review, 73(2), 89-95.
\item Singh, M., \& Kim, S. (2019). Blockchain technology for decentralized
autonomous organizations. In Advances in computers (Vol. 115, pp.
115-140). Elsevier.
\item Sipser, M. (2006). Introduction to the Theory of Computation (Vol.
2). Boston: Thomson Course Technology.
\item Slamka, C., Skiera, B., \& Spann, M. (2012). Prediction market performance
and market liquidity: A comparison of automated market makers. IEEE
Transactions on Engineering Management, 60(1), 169-185. 
\item Sneed, H. M., \& Merey, A. (1985). Automated software quality assurance.
IEEE Transactions on Software Engineering, (9), 909-916.
\item Srivastava, S. (2023). Prospective Application of Blockchain in Mutual
Fund Industry. In New Horizons for Industry 4.0 in Modern Business
(pp. 121-148). Cham: Springer International Publishing.
\item Stulz, R. M. (2007). Hedge funds: Past, present, and future. Journal
of Economic Perspectives, 21(2), 175-194.
\item Syed, T. A., Alzahrani, A., Jan, S., Siddiqui, M. S., Nadeem, A.,
\& Alghamdi, T. (2019). A comparative analysis of blockchain architecture
and its applications: Problems and recommendations. IEEE access, 7,
176838-176869.
\item Tapscott, D., \& Tapscott, A. (2016). Blockchain revolution: how the
technology behind bitcoin is changing money, business, and the world.
Penguin.
\item Thakkar, P., Nathan, S., \& Viswanathan, B. (2018, September). Performance
benchmarking and optimizing hyperledger fabric blockchain platform.
In 2018 IEEE 26th international symposium on modeling, analysis, and
simulation of computer and telecommunication systems (MASCOTS) (pp.
264-276). IEEE.
\item Tikhomirov, S. (2018). Ethereum: state of knowledge and research perspectives.
In Foundations and Practice of Security: 10th International Symposium,
FPS 2017, Nancy, France, October 23-25, 2017, Revised Selected Papers
10 (pp. 206-221). Springer International Publishing.
\item Tolmach, P., Li, Y., Lin, S. W., Liu, Y., \& Li, Z. (2021). A survey
of smart contract formal specification and verification. ACM Computing
Surveys (CSUR), 54(7), 1-38.
\item Tufano, P. (2003). Financial innovation. Handbook of the Economics
of Finance, 1, 307-335.
\item Vacca, A., Di Sorbo, A., Visaggio, C. A., \& Canfora, G. (2021). A
systematic literature review of blockchain and smart contract development:
Techniques, tools, and open challenges. Journal of Systems and Software,
174, 110891.
\item Wang, S., Yuan, Y., Wang, X., Li, J., Qin, R., \& Wang, F. Y. (2018).
An overview of smart contract: architecture, applications, and future
trends. In 2018 IEEE Intelligent Vehicles Symposium (IV) (pp. 108-113).
IEEE.
\item Wang, S., Ding, W., Li, J., Yuan, Y., Ouyang, L., \& Wang, F. Y. (2019).
Decentralized autonomous organizations: Concept, model, and applications.
IEEE Transactions on Computational Social Systems, 6(5), 870-878.
\item Werner, S. M., Perez, D., Gudgeon, L., Klages-Mundt, A., Harz, D.,
\& Knottenbelt, W. J. (2021). Sok: Decentralized finance (defi). arXiv
preprint arXiv:2101.08778.
\item Wood, G. (2014). Ethereum: A secure decentralised generalised transaction
ledger. Ethereum project yellow paper, 151(2014), 1-32.
\item Wu, H., Cao, J., Yang, Y., Tung, C. L., Jiang, S., Tang, B., ... \&
Deng, Y. (2019, July). Data management in supply chain using blockchain:
Challenges and a case study. In 2019 28th International Conference
on Computer Communication and Networks (ICCCN) (pp. 1-8). IEEE.
\item Wu, S. X., Wu, Z., Chen, S., Li, G., \& Zhang, S. (2021). Community
detection in blockchain social networks. Journal of Communications
and Information Networks, 6(1), 59-71.
\item Zeng, X., Hao, N., Zheng, J., \& Xu, X. (2019). A consortium blockchain
paradigm on hyperledger-based peer-to-peer lending system. China Communications,
16(8), 38-50.
\item Zetzsche, D. A., Arner, D. W., \& Buckley, R. P. (2020). Decentralized
finance. Journal of Financial Regulation, 6(2), 172-203.
\item Zheng, Z., Xie, S., Dai, H., Chen, X., \& Wang, H. (2017, June). An
overview of blockchain technology: Architecture, consensus, and future
trends. In 2017 IEEE international congress on big data (BigData congress)
(pp. 557-564). Ieee.
\item Zheng, Z., Xie, S., Dai, H. N., Chen, W., Chen, X., Weng, J., \& Imran,
M. (2020). An overview on smart contracts: Challenges, advances and
platforms. Future Generation Computer Systems, 105, 475-491.
\item Zheng, Y., \& Boh, W. F. (2021). Value drivers of blockchain technology:
A case study of blockchain-enabled online community. Telematics and
Informatics, 58, 101563.
\item Zou, W., Lo, D., Kochhar, P. S., Le, X. B. D., Xia, X., Feng, Y.,
... \& Xu, B. (2019). Smart contract development: Challenges and opportunities.
IEEE Transactions on Software Engineering, 47(10), 2084-2106.
\end{itemize}

\end{document}